\documentstyle[12pt,aasms4]{article}
\begin{document}
\renewcommand\floatpagefraction{.9}
\renewcommand\topfraction{1}
\renewcommand\bottomfraction{1}
\renewcommand\floatsep{12pt}
\renewcommand\textfraction{0}
\renewcommand\intextsep{12pt}

\title{ HUBBLE SPACE TELESCOPE IMAGES OF A SAMPLE OF 
TWENTY NEARBY LUMINOUS QUASARS
\footnote{Based on observations with the NASA/ESA Hubble Space
Telescope, obtained at the Space Telescope Science Institute, which is
operated by the Association of Universities for Research in Astronomy,
Inc., under NASA contract NAS5-26555.\smallskip}}
\author{John N. Bahcall, Sofia Kirhakos, and David H. Saxe,}
\affil{Institute for Advanced Study, School of Natural Sciences,
Princeton, NJ~08540}
\centerline{and}
\author{Donald P. Schneider}
\affil{Department of Astronomy and Astrophysics, The Pennsylvania 
State University, University Park, PA 16802}

\begin{abstract}

Observations with the Wide-Field Camera of the {\it Hubble Space
Telescope (HST)} are presented for a representative sample of 20
intrinsically luminous quasars with redshifts smaller than 0.30. These
observations show that luminous quasars occur in diverse environments
that include ellipticals as bright as the brightest cluster galaxies
(2), apparently normal ellipticals (10), apparently normal spirals
with H~II regions (3), complex systems of gravitationally interacting
components (3), and faint surrounding nebulosity (2). 

The quasar host
galaxies are centered on the quasar to the accuracy of our
measurements, 400~pc.  
Some of the host galaxies show no evidence of merging or strong
gravitational interactions.
There are more radio quiet quasars
in galaxies that appear to be ellipticals (7) than in spiral hosts
(3), contrary to expectations. However, three, and possibly five, of
the six radio loud quasars have detectable elliptical hosts, in
agreement with expectations.  
Strong upper limits are placed on the possible existence of optical
jets.
The luminous quasars studied in this paper
occur preferentially in luminous galaxies.
The average  absolute magnitude  of the hosts is $2.2$ 
magnitudes brighter than expected for a field galaxy luminosity
function.

The superb optical characteristics of the repaired $HST$ make possible
the detection of close galactic companions; we detect eight companion
galaxies within projected distances of 10~kpc from the quasars.  The
presence of very close companions, the images of current gravitational
interactions, and the higher density of galaxies around the quasars
suggest that gravitational interactions play an important role in
triggering the quasar phenomenon.

\end{abstract}

\section{INTRODUCTION}
\label{introduction}

Figures~\ref{figallqso} and \ref{figallqsosub} tell 
the main story of this paper. We urge the reader
to look at these beautiful $HST$ images before continuing with the
text and the quantitative details.

We summarize in this paper the results of our analysis of $HST$-WFPC2
observations of a representative sample of 20 of the most luminous 
(M$_{\rm V}< -22.9$) nearby ($z < 0.30$) quasars. The goal of these
observations was to help understand the quasar phenomenon by
determining the environment in which quasars occur. The main result of
this paper is that there is not one type of environment, but instead a
wide range of environments in which the most luminous quasars appear
to be embedded. The $HST$ images also contain a number of
extraordinary phenomena and some surprises, including: host galaxies
that appear normal and show no evidence of strong gravitational
interactions, very close galactic 
companions, host ellipticals for radio quiet quasars, spiral hosts
with well developed arms and prominent H~II regions, galaxies caught in
the act of merging, apparently faint galactic hosts, and very
extended emission.  

Some partial results of this study have been
reported previously: Bahcall et al.\ (1994, 1995, 1995a,b,c, 1996).
For those aspects of the work that depend upon the subtraction of a
stellar point spread function (PSF), there are small, quantitative 
 differences between the results 
described in this paper and previous results we have reported.
In the previous work, we used a stellar PSF determined from a red
standard star, F141. 
In the present work, we have used stellar PSFs that were obtained for
four separate blue stars (see discussion in Section~\ref{aftersub} of
the PSFs constructed by Krist and Burrows 1996  ).
The visual appearance of the hosts in the 
subtracted images is, in a few cases, 
significantly improved by using the PSFs of the blue stars.

Other $HST$ studies of quasar imaging (although many of the objects do
not satisfy our luminosity criterion) include Hutchings \& Morris
(1995), Hutchings et al., (1994), and 
Disney et al.\ (1995).

The subject of quasar environments has a long and distinguished
history. Some representative papers that report on ground-based
observations are: Kristian~(1973), Wyckoff et al. (1980), Wyckoff,
Wehinger, \& Gehren (1981), Tyson, Baum, and Kreidl (1982) Hutchings
et al. (1982), Gehren, et al.~(1984), Heckman et al.~(1984), Malkan
(1984), Malkan, Margon, \& Chanan (1984), Boroson, Persson, \&
Oke~(1985), Smith et al.~(1986), Hutchings (1987), Stockton \&
MacKenty~(1987), Yee~(1987), Hutchings, Janson, \& Neff~(1989),
Romanishin \& Hintzen (1989), V\'eron-Cetty \& Woltjer (1990),
Hutchings \& Neff (1992), Dunlop et al.~(1993), and McLeod \& Rieke
(1994a,b).  

The analyses described in these pioneering ground-based studies are
made  difficult because of atmospheric seeing; the
light from the bright central (nuclear) sources may be a few
magnitudes brighter than the total emission from the host galaxies.
Nevertheless, there is agreement (within typically one magnitude or
better) between our $HST$ observations and previous ground-based
estimates of the total apparent magnitudes of the host emission,
although $HST$ reveals details not previously accessible and corrects
some important conjectures that are not supported by the
higher-resolution observations.

The paper is organized as follows. In Section~\ref{obssection} we
describe the sample selection and observations; in
Section~\ref{subunproc} we present the unprocessed data; in
Section~\ref{aftersub} we describe the method of removing the light
due to the quasar (point spread function subtraction) and present the
data after a stellar PSF is subtracted; in Section~\ref{anasection} we
describe the methods of analysis of the data; in Section~\ref{secsize}
we report on the measurements of the host galaxies; in
Section~\ref{companion} we discuss the presence of companion galaxies;
in Section~\ref{groundsection} we compare our measurements with
results from some ground-based observations; in Section~\ref{comments}
we comment on the environment and host galaxy of each quasar, and in
Section~\ref{discuss} we summarize and discuss our results.  We assume
in this paper that $H_0~=~100~ {\rm km~s^{-1}Mpc^{-1} }$ and
$\Omega_0~=~1.0$.

\section{SAMPLE SELECTION AND OBSERVATIONS}
\label{obssection}

We describe in this section how the sample was selected 
(\S~\ref{subsample}) and how 
the observations were performed (\S~\ref{subobs}).

\subsection{Sample Selection}
\label{subsample}

A sample of 14 quasars was selected solely on the basis 
of luminosity ($M_V<-22.9$), redshift ($z \le 0.20$), and 
galactic latitude
($\vert b
\vert > 35^\circ$). All the quasars in the V\'eron-Cetty \&
V\'eron~(1991) catalog that satisfied the redshift,  luminosity, and
galactic latitude 
criteria were included in this  sample; no distinction was
made on the basis of radio or other secondary properties.  
 One may choose to call this set
of quasars a ``complete sample'' within the context of the 
V\'eron-Cetty \&
V\'eron~(1991) catalog\footnote{The concept of a ``complete'' sample of
quasars requires clarification since a variety of techniques,
including radio emission, optical colors, variability, and x-ray
emission, are used to discover quasars.  It is conceivable, indeed
likely, that there is at least one object satisfying our defining
sample criteria that has not yet(in 1996) been recognized
observationally.  
For lack of a better term, we use here the designation ``complete sample''
in a limited sense
to mean all the objects within a specified catalog having  stated
characteristics.}.

The 14 quasars with $z\le 0.20$ have an average (median) 
absolute magnitude $\langle M_V \rangle = -23.4\ (23.2)$ and an average
redshift $\langle z \rangle = 0.17$.
Only 3
radio loud quasars are  present in the original  sample of 14
objects. 
In the V\'eron-Cetty and
V\'eron (1996) catalog, there are an additional 4 quasars with $z \le
0.20$ that satisfy our luminosity and galactic latitudes requirements;
all are radio quiet.

By combining the time available from GTO and GO programs, an
additional 6 quasars  with redshift between $0.20 < z <
0.30$ were added to the original sample of 14 objects; 
these additional objects satisfied the same luminosity and
galactic latitude constraints as the original  sample. 
The additional objects contain 3 radio loud quasars; 
the total sample of 20 contains 6 quasars that are classified as
radio loud, i.e.  with $L_{\rm 5GHz}
\gtrsim {\rm 10^{26}~W~Hz^{-1}}$  (Kellermann et al.\ 1994).  
 The additional 6 objects are
slightly brighter in the optical on the average (median), 
$\langle M_V \rangle = -24.0\ (-24.1)$, and
have slightly larger redshifts, 
$\langle z \rangle = 0.26$.  The range of absolute magnitudes in
the 6 added objects, $-24.6 \le M_V < -22.9$, is included within the
range of absolute magnitudes spanned by the original sample of 14
objects, i.e., $-25.6 \le M_V < -22.9$.

In what follows, we shall refer to the total sample of 20 quasars as a
``representative sample'' with $M_V < -22.9$ and $z < 0.30$.  The
average (median) absolute magnitude for the full sample is $-23.6\
(-23.2)$; the average redshift $\langle z\rangle = 0.19$.
Nearly all (18) of
the quasars have $15.1 < V < 16.7$, but two are much brighter in the
optical band: 3C~273 ($V=12.8$) and HE~1029$-$140 ($V=13.9$).

\subsection{Observations}
\label{subobs}

A journal of the observations is given in Table~\ref{journal}, which
lists the following quantities for each object: the date observed, the
longest exposure time of a single image, the detected number of ${\rm
electrons~pixel^{-1}~sec^{-1}}$ in the sky, the quasar redshift, the
distance in kiloparsecs that corresponds to an angular separation of
1\arcsec\ as seen from Earth, the apparent $V$-magnitude (from
V\'eron-Cetty \& V\'eron 1996), the absolute $V$-magnitude (without
k-correction), the radio properties (an ``X'' identifies the radio
loud quasars), and in the last column the existence of $HST$
spectroscopy (a ``K'' indicates that the $HST$/FOS observations were
taken as part of the $HST$ Quasar Absorption Line Key Project, and
some of the results are reported by Bahcall et al.\ (1993) and Jannuzi
et al.\ (1997); an ``O'' indicates that other FOS observations exist).

The quasars were observed with the Wide Field/Planetary Camera-2
(WFPC2) through the F606W filter, which is similar to the $V$~bandpass
but is slightly redder; the mean wavelength and FWHM of the F606W
system response are 5935~\AA\ and 1500~\AA, respectively.  The F606W
filter was chosen because of its high throughput. In all of the
quasars discussed here, redshifted $H\beta$ and [O~III] are included
in the bandpass.  At a given angular radius, the scattered light in
the Wide Field Camera-2 (WF) is about five times less than the scattered
light in the Planetary Camera (see Krist \& Burrows 1994).  We
chose to use the WF instead of the PC in the original formulation of
this program because of the likelihood that the host galaxies would
have low surface brightnesses that extended over areas large compared
to the WF resolution ($0.1''$ or about 0.2 kpc). The results
reported here support the original choices, since the observed galaxy
extensions are indeed large compared to the WF resolution.

The center of light of all quasars was placed  within $4\arcsec \pm
1\arcsec$\ from the center of the Wide-Field Camera CCD~3 (WF3). Three
exposures were taken of each quasar. The integration times for 14
objects were 1400~s, 500~s, and 200~s; the exposures for the remaining
six objects were 1100~s, 600~s, and 100~s. 

  The size of WF3 is $800 \times 800$ pixels (exposed area $\sim 770 \times
750$ pixels) and its image scale is 0\farcs0996\ pixel$^{-1}$.  We
report measured F606W magnitudes on the $HST$ photometric scale
established by Holtzman et~al.~(1995b). The adopted photometric
zero-point for 1~${\rm electron~sec^{-1}}$ is 24.94~mag for the F606W filter.  
For
further information about the WFPC2, see Burrows~(1994), Trauger
et~al.~(1994), and Holtzman et~al.~(1995a,b).  Additional details of
the observational procedures are given in Paper~I and Paper~II.

The innermost regions of the quasar images are saturated in all of the
longest exposures out to a radius $\approx 0.3''$, except for the the two
optically brightest quasars in our sample, HE~1029$-$140 and 3C~273,
in which cases the images were saturated out to $\approx 0.5''$ and
$0.7''$, respectively.  The number of saturated pixels in the central
region of the quasar images were typically 30 pixels for the longest
exposures. In addition, $\sim 15$ saturated pixels were present due to
the ``vertical bleeding''.  For HE~1029$-$140 and 3C~273, the number
of saturated pixels was approximately 105 and 190, respectively, plus
90 and 390 from the vertical bleeding.

The initial data processing (bias frame removal and flat-field
calibration) was performed at the Space Telescope Science Institute
with their standard software package.  The individual images of each
quasar were aligned to better than 0.3~pixel; this made it easy to
identify and eliminate cosmic ray events.  Cosmic rays were identified
by a pixel-by-pixel comparison of pairs of images; the intensity of a
pixel containing a cosmic ray was replaced by the scaled value of the
intensity of the pixel in the other image.  The flat-field corrections
were based upon pre-flight calibrations; these calibrations remove the
small-scale (few pixel) sensitivity variations.  The typical signal
and rms of the noise of the sky in the long exposures (in detected
photons per pixel) are 157 and 14, respectively.  

The sky level
corresponds to an average surface brightness of approximately
$22.2$~mag~arcsec$^{-2}$. The limiting surface brightness at which
objects can be detected is
typically between $25$ and $26~{\rm mag~arcsec^{-2}}$ (F606W), cf. 
Table~\ref{size}.

\section{THE UNPROCESSED IMAGES}
\label{subunproc}

We present in this section the images as received from STScI, without
further processing except for the removal of cosmic rays.

Figure~\ref{figallqso} shows a $23\arcsec \times 23\arcsec$\ WF image
of each of the 20 quasars, plus the image of a blue star (upper
left-hand portion). The images displayed are the longest individual
exposures we have.  Many, but not all, of the quasars are noticeably
non-stellar, and host galaxies are visible on the unprocessed $HST$
images.  The exposure time of the star image shown in the top panel of
Figure~\ref{figallqso} is 20~s, which yields a total number of counts
similar to that obtained in the quasar images.  The star is MMJ~6490
in the M67 cluster.  Its apparent $V$ magnitude is 10.99 and its $B-V
= 0.11$ (Montgomery, Marschall \& Janes 1993).

Some features of host galaxies are obvious in
Figure~\ref{figallqso}. Normal spiral galaxies, with prominent H~II
regions, envelop PG~0052$+$251 and PG~1402$+$261. Host elliptical
galaxies are clearly seen in the images of PHL~909, HE~1029$-$140,
PG~1116$+$215, and 3C~273. There are three obvious cases of current
gravitational interaction: 0316$-$346, PG~1012$+$008, and
PKS~2349$-$014.

Some of the host galaxies are seen even in our shortest exposures
(200~s, see Bahcall et al.\ 1996 for short exposure images of
PG~0052$+$251 and PHL~909).  In some cases, like NAB~0205$+$02,
PG~0953$+$414, and PG~1307$+$085, it is difficult to distinguish the
quasar image from the star even in our longest exposures.

\section{IMAGES AFTER SUBTRACTION OF A STELLAR PSF}
\label{aftersub}
 
    The major challenge in the analysis of these $HST$ data is the
removal of the light produced by the quasar, which is presumed to be a
point source.  We have adopted an empirical approach.  We have used
images of stars observed at the same location on the detector as the
quasars to determine the point-spread function, PSF, of the stellar
quasar.

For about half of the quasars considered here, the principal
observational results concerning the host environment can be obtained
without the PSF subtraction.  Examples of exceptions are the hosts of
NAB~0205$+$02, PG~0953+414, PG~1444$+$407, and 3C~323.1, which are
more apparent in Figure~\ref{figallqsosub} than in
Figure~\ref{figallqso}.  Close companions for PKS~1302$-$102,
PKS~2135$-$147, and PKS~2349$-$014 are also more clearly visible after
the subtraction of a stellar PSF.

The stellar PSF was measured in Cycle~5 (HST program 5849) by obtaining
a set of 13 images for each of four blue stars in the
M67 cluster. The calibration stars (and their $B-V$ colors) are:
MMJ~6481($-0.073$), MMJ~6490($0.11$), MMJ~6504($0.22$), and
MMJ~6511($0.34$).
The apparent magnitudes range from $V = 10.03$ to $V = 10.99$.
For each star, a series of four images were used 
by Krist and Burrows (1996) to produce a PSF
that samples the full dynamic range of the star image and covers the
saturation range found in the quasar images.
In these calibration
images, the radius of the saturated region varies from $0.0''$ to
$0.6''$; the exposure times range from $0.1$~s to $100$~s. 

The PSF data are publicly available at 
http://www.sns.ias.edu/\raise2pt\hbox{${\scriptstyle\sim}$}jnb. 
There is detailed documentation regarding 
the  PSF data and their use at this site (go to HST Images)
by J. Krist and C. Burrows describing
how the PSFs were constructed.

Whenever PSF subtractions were required, we used all four PSFs
determined by Krist and Burrows.  For each case, we selected the
result that gave the cleanest subtraction (i.e., the fewest artifacts
produced).

We fit a stellar point-spread function to each quasar image and
 subtracted a multiple of the normalized PSF to search for underlying
 diffuse light from hosts.  The best fit was determined by minimizing
 the differences between the quasar and the PSF using a
 $\chi^2$-routine calculated in two distinct areas: azimuthal averages
 and narrow regions centered on the diffraction spikes.  The two
 methods gave essentially same results (see Paper~II), differing in
 inferred host galaxy magnitude by $\pm 0.1$~mag.  The quasar images
 with the PSF subtracted presented in this paper were obtained by
 minimizing the $\chi^2$ in azimuthal averages.  
%Sofia: Pls. check wording of above sentence.

We have estimated some of the likely systematic uncertainties 
in the subtraction process by
subtracting a best-fit PSF of the standard star, MMJ~6504, from 
the image of another standard 
star, MMJ~6490. We have Cycle~5 observations of
MMJ~6490 observed at the same position in WF3 as the sample quasars.
Figure~\ref{figallqsosub}a (first panel) shows 
the image of the best fit of MMJ~6504 subtracted from MMJ~6490.
The PSF subtraction, star 
from star, is very good, although some faint diffuse residuals are
still present. If we scale the intensity of MMJ~6490
so that the central brightness corresponds to the apparent magnitude
of one of our typical quasars, i. e., $V=15.6$, then the residual
``nebulosity'' left over when the MMJ~6504  PSF is subtracted is about
21.0~mag.

Figure~\ref{figallqsosub} shows the images for all the quasars in our
sample after a best-fit stellar point-spread function (PSF) 
was subtracted from the original
images shown in Figure~\ref{figallqso}. 

Table~\ref{size} lists the major diameter of the host galaxies in
arcseconds and in kpc, the surface brightness of the isophote for
which the size was measured, and a tentative morphological
classification.  For each quasar, the magnitude of the host galaxy was
calculated using three different methods, aperture photometry (see
\S~\ref{subap}), and one-dimensional and two-dimensional galaxy
modeling (see \S~\ref{subrp} and \S~\ref{subtwo}).

We determined the size of the quasar hosts by examining the PSF
subtracted images and measuring the faintest isophotes of
the host galaxy. The morphological classification was done by visual
inspection of the images and following as closely as possible the
classification given in {\it The Hubble Atlas of Galaxies} by Sandage
(1961).  In a number of cases indicated by (?) in Table~\ref{size}, we
have denoted as ``En'' featureless, smooth hosts that do not show any
obvious discontinuities or morphological features.

\section{METHODS OF ANALYSIS}
\label{anasection}

In this section, we describe the different methods of analysis that we
have used to determine the properties of the quasar hosts. The
measured quantities are presented in Tables~\ref{size}--\ref{veron}
and discussed in \S\S~\ref{secsize}--\ref{companion}. We describe in
 \S\S~\ref{subap} to
\ref{subtwo}  different methods for estimating the
magnitude of the host galaxies.  We discuss in \S~\ref{subap} our
results for aperture photometry, 
in \S~\ref{subrp} the radial profile fits, and in
\S~\ref{subtwo} the two-dimensional fits.

Some of the host environments are complex, including merging or
tidally interacting galaxies and close companions.
The smooth de Vaucouleurs or exponential disk models used in
\S~\ref{subrp} and \ref{subtwo} are obviously not realistic
descriptions of the light distribution for complex environments. In
order to provide a common basis of comparison with ground based
observations, we provide the results of fits with smooth models for
all of the host environments, complex or not.

\subsection{Aperture Photometry}
\label{subap}

We performed aperture photometry in circular annuli centered on the
quasars, after a best-fit PSF was subtracted. An inner radius of
$r=1.0\arcsec$\ was used for all quasars, except for HE~1029$-$140 and
3C~273.  In most observations, the region $r<1\arcsec$\ is
contaminated by artifacts left by the PSF subtraction. The saturated
areas in the images of HE~1029$-$140 and 3C~273 are larger; for those
two cases, the inner radii used were 1.5\arcsec\ and 2.0\arcsec,
respectively.  The outer radii chosen in general  represent how far we
could see the host galaxy (see Table~\ref{result}).
  The aperture magnitudes calculated using
an inner radius $r=1.0\arcsec$\ are typically $\sim 0.6$~mag fainter
than the total magnitude for the galaxy that was estimated by fitting
a model (see \S\S~\ref{subrp}--\ref{subtwo}) to the measured surface
brightness. A difference between aperture and model magnitudes is
expected because the aperture magnitudes do not include the area
within 1\arcsec\ of the quasar; all of the models we considered have
surface brightnesses that increase monotonically toward the center.

\subsection{One-Dimensional Radial Profiles}
\label{subrp}

The one-dimensional azimuthally-averaged surface brightness profiles
of the host galaxies were constructed from the $HST$ data after
subtraction of a best-fit stellar PSF.  Regions affected by
saturation, diffraction spikes, or residual artifacts from the PSF
subtraction were not included in the azimuthally averaged
profiles. For each galaxy, we obtained a best-fitting exponential disk
(henceforth Disk) and a de Vaucouleurs (1948, henceforth GdV) profile
that fits the observed data in the region  $r\ge 1\arcsec$. The total
magnitudes obtained this way are systematically brighter than the ones
obtained by aperture photometry (\S~\ref{subap}) which excludes the
innermost, saturated regions of the profiles.

\subsection{Two-Dimensional Fit}
\label{subtwo}

The $HST$ imaging provides greater detail than has been available
previously in ground-based images of luminous quasars.
Traditionally, the properties of host galaxies have been determined in
ground-based studies by making model fits to azimuthally-averaged
radial profiles (see \S~\ref{subrp}).

To take advantage of the $HST$ resolution, we have developed software to
fit a two-dimensional model to the PSF-subtracted quasar images. For
each quasar, we fit an analytic galaxy model (exponential disk or de
Vaucouleurs profile) to the data and calculated the $\chi^2$.  The
area used for the fit was approximately an annular region, centered on
the quasar, that excluded the central area ($r < 1.0$ \arcsec), and
the remnants of the diffraction spikes or other artifacts clearly due
to improper PSF subtraction.  We fit four parameters: the (x,y) pixel
position of the center, the total number of counts, and the radius
(effective radius or scale length) in the galaxy model.

We begin the iteration by giving the
software the position of the quasar nucleus, and calculate only the
total number of counts.  The number of counts found in this step is
 entered as the initial guess for the galaxy brightness, and the
program then fits the counts and radius, keeping the position fixed.
Finally, we supply the software with the previously calculated counts
and radius and ask the software to fit all four parameters.  

We tested the software for the two-dimensional fits in different
ways.  Initially we created two model galaxies: a disk and an
elliptical galaxy. We checked how well the two-dimensional software
reproduced the position of the center of the model galaxies, the scale
length or effective radius, and the total number of counts. We made
extensive checks by varying the input parameters and the order in which
the parameters were
fit. The best results were achieved when we followed the three step
iterative process in the order described above. We fit each model galaxy
with an exponential disk and a de Vaucouleurs model - the smallest
$\chi ^2$ residuals were obtained when the disk galaxy was fit by an
exponential disk model, and the elliptical was fit by a de Vaucouleurs
model.  For the model galaxies, the discrepancy in position and size
were $< 1$ pixel. The discrepancy in the total number of counts was
$\sim 1\%$. The accuracy achieved in the simulations is greater than
can be achieved with the real data, because of the imperfect
subtraction of the PSF for the $HST$ images. 

The host galaxies of PG~0052$+$251 and PHL~909 were also used to test
the two-dimensional software. We compared the software output position
with our measured position for the center of the galaxy (agreement
better than 0.2\arcsec), compared the total number of counts with the
value we estimated from aperture photometry (agreement turned out to
be 0.1~mag for PG~0052+251, and 0.5~mag for PHL~909), and verified
that the output scale length or effective radius were plausible.

\section{MAGNITUDES AND POSITIONS OF HOST GALAXIES}
\label{secsize}

In this section, we report on the measurements of the 
magnitude and position of the host galaxies.

In Table~\ref{result} we list the magnitudes for the host galaxies
measured using aperture photometry and for a few cases the surface
brightness method. The inner and outer radii used in performing
aperture photometry are listed as well. The outer radius indicates how
far from the quasar the host galaxy was clearly visible. We
transformed the measured F606W aperture magnitudes to $V$ applying
k-corrections calculated by Fukugita et al.\ (1995). Given the redshift
of the quasar and the morphological type of the host, we used the
Fukugita et al. Table~6 and Figure~14b to obtain the ($F606W - V$)
color.  For the cases in which we are uncertain about the
morphological type of the host, we assumed an average of the ($F606W -
V$) color for ellipticals and Sab galaxies. 

Table~\ref{model} lists the results for the one-dimensional and
two-dimensional galaxy model fitting. The two-dimensional models are
somewhat fainter than the corresponding one dimensional fits.
Specially, we find:
\begin{equation}
\langle m_{\rm 606W,2-D}- m_{\rm 606W,1-D} \rangle = \left\{\matrix{0.4 \pm
0.2,&\rm  GdV\cr
0.3 \pm 0.1,&\rm Disk\cr}\right.\ .
\label{eq:1}
\end{equation}

The magnitudes obtained by model fits are brighter than those obtained
by aperture photometry since the models include estimated
contributions from the inner (saturated) regions of the images.  On
average
\begin{equation}
\langle m_{\rm 606W,1-D}- m_{\rm 606W,aperture} \rangle 
= \left\{\matrix{-1.1 \pm 0.2,&\rm GdV\cr
- -0.4 \pm 0.2,&\rm Disk\cr}\right.\ .
\label{eq:2}
\end{equation}
As equation (3) shows, the magnitudes of the host
galaxies estimated by fitting a GdV model to the azimuthal averaged
radial profile of the residual light are on average 1.0~mag brighter
than the magnitudes obtained from aperture photometry; fitting an 
exponential disk model gives magnitudes that are on average 0.5
brighter than the results from aperture photometry.
For the two-dimensional model fits, we have on average
\begin{equation}
\langle~\vert~ m_{\rm 606W,2-D}- m_{\rm 606W,aperture}~\vert~\rangle 
= \left\{\matrix{-0.7 \pm 0.2,&\rm GdV\cr
- -0.2 \pm 0.1,&\rm Disk\cr}\right.\ .
\label{eq:3}
\end{equation}

Table~\ref{bestmodel} lists our best-estimate absolute $V$ magnitudes.
In computing the entries in Table~\ref{bestmodel}, we used the
k-correction of Fukugita et al.\ (1995) that are listed in
 the next-to-last column of 
Table~\ref{result}. In computing the absolute magnitudes, 
we selected the best-fit model based on the
morphology of the host galaxy, unless the morphology is uncertain, and
for  those cases we list the model that gives the smallest $\chi ^2$
residuals.

The
results from the two-dimensional fits indicate that the host galaxies
are, on average, centered on the quasar with:

\begin{equation} 
\langle~{\Delta r}~\rangle ~=~ 0.4 \pm 0.4~{\rm kpc}.
\label{eq:4}
\end{equation}
The host galaxies are typically centered within 400~pc of the location
of the quasar point source. If we eliminate from the comparison the
most extreme examples of especially complex environments (0316$-$346, 
PG~1012$+$008, and PKS~2349$-$014) the results are,

\begin{equation} 
\langle~{\Delta r}~\rangle ~=~ 0.2 \pm 0.2~{\rm kpc}.
\label{eq:5}
\end{equation}

The 2-D (1-D) GdV model gives magnitudes for the host that are on average
0.6~mag (0.7~mag) brighter than the exponential disk estimates.

\section{COMPANION GALAXIES}
\label{companion}

Inspection of the $HST$ images of the quasar fields reveals a number
of companion galaxies projected close to the quasars.  The relatively
long exposures (1100 or 1400 seconds), combined with the excellent
angular resolution, allowed galaxies to be identified down to limiting
magnitude $m(F606W) \lesssim 25.0$ and as close as 1\arcsec\ or 2\arcsec\
from the central quasar. We performed aperture photometry on the
galaxies in the quasar fields
using circular apertures with radii of $0.3''$ to $10''$, as
appropriate.

We counted the number of companion galaxies brighter than a specified
limiting absolute magnitude that were found to have a metric
separation from one of the quasars of less than or equal to some predetermined
distance.  We choose {\it a priori} a limiting absolute magnitude of
$M_V = -16.5$ (four magnitudes fainter than $L^*$) and a maximum
separation of 25 kpc (see Paper~II).  

Table~\ref{comps} lists all galaxies found around the quasars that
satisfy these specifications.  This table gives for each quasar the
number of companion galaxies that are at least as bright as $M_V =
- -16.5$ (if they have the same redshift as the quasar) 
and that are projected within 25 kpc of the center of light of
the quasar.  The separations both in arcseconds and in kpc are also
given in Table~\ref{comps}, along with the brightnesses of the companion
galaxies, tabulated in both apparent and absolute magnitude.

The density of companion galaxies brighter than $M_V=-16.5$, within
25~kpc of the quasars, may be 
higher around quasars with elliptical hosts. There are 
 2 companions for 4
spiral hosts  and  13 companions for 12 elliptical hosts.

Fisher et al.\ (1996) examined the clustering of galaxies around
all the quasars in this sample and found  a significant enhancement of
galaxies within a projected separation of $\lesssim 100$~$ {h^{-1}\rm kpc}$\ 
of the
quasars.  Modeling the quasar/galaxy correlation function as a power
law with a slope given by the galaxy/galaxy correlation function,
Fisher et al.\
find that the ratio of the quasar/galaxy to galaxy/galaxy correlation
functions is $3.8\pm 0.8$.  The galaxy counts within $r<15$~$ {h^{-1}\rm 
kpc}$\ of
the quasars are too high for the density profile to have an
appreciable core radius ($\gtrsim 100$~$ {h^{-1}\rm kpc}$).  These results 
provide
further support for 
the idea that low redshift quasars are located preferentially in
groups of 10--20 galaxies rather than in rich clusters.  Fisher et al.\
do not detect a 
significant difference in the clustering amplitudes derived from
radio loud and radio quiet subsamples.

\section{COMPARISON WITH GROUND-BASED OBSERVATIONS}
\label{groundsection}

In this section we compare the results of $HST$-based images of
luminous quasars with previously-published representative analyses of 
ground-based
observations. In \S~\ref{subsecveron}, we compare our $HST$
measurements with V\'eron-Cetty \& Woltjer (1990) $i$-band results, in
\S~\ref{subsecdunlop}, with Dunlop et al.\ near-infrared images,
and in \S~\ref{subsecmcleod}, with McLeod \& Rieke (1994b)
$H$-band observations.
 
\subsection{V\'eron-Cetty \& Woltjer: Annular Regions}
\label{subsecveron}

V\'eron-Cetty \& Woltjer (1990) suggested that the apparent magnitudes
of host galaxies should be measured in a fixed metric annulus that is
well removed from the quasar nucleus.  They proposed an annular region
of $12.5$ kpc to $25.0$ kpc for $\Omega_0 = 0.0$ and $H_0 = 50$ km
s$^{-1}$Mpc$^{-1}$.  The V\'eron-Cetty \& Woltjer proposal compares
directly measurements of the same quantity made by separate groups
using 
different techniques. In this way, measurement uncertainties can
be separated from differences caused by the variety of choices in the
models used to fit to the observations.  

We have three objects in common with V\'eron-Cetty \& Woltjer:
PKS~1302$-$102, PKS~2135$-$147, and PKS~2349$-$014. The specified
annular regions are 2.17\arcsec\ and 4.35\arcsec, 2.82\arcsec\ and
5.63\arcsec, and 3.15\arcsec\ and 6.30\arcsec\ for the three quasars,
respectively.  All three quasars have close companion galaxies in the
regions specified by V\'eron-Cetty \& Woltjer (see
Table~\ref{comps}). The close companions of PKS~1302$-$102 and
PKS~2349$-$014 were not noticed on the ground-based images.

Table~\ref{veron} summarizes the aperture photometry that was
performed in the same annular regions as V\'eron-Cetty \& Woltjer. We
list the $i$-band annular magnitudes obtained by V\'eron-Cetty \&
Woltjer and their estimated absolute $V$-magnitude, converted to the
cosmological parameters used in this paper, the absolute magnitude we
measured in the $HST$ images (excluding the light of the companions)
with the F606W, and our estimated absolute $V$-magnitude. We also
include in the table the total absolute $V$-magnitude V\'eron-Cetty \&
Woltjer obtained for the host galaxies (fitting a spheroidal model)
and our estimated 2-D model $V-$band host galaxy magnitude.

The agreement between  our results and the ground-based observations
of V\'eron-Cetty \& Woltjer for the aperture
photometry between 12.5 kpc and 25 kpc is satisfactory, but not as
precise as we would have hoped. 
The average difference between our estimated $M_V$ and that of
V\'eron-Cetty \& Woltjer is

\begin{equation} 
{\langle{M_{V(i)} - 
M_{V(F606)}}\rangle}_{(12.5-25{\rm kpc})VCW}\  =\  -0.4 \pm 0.1~{\rm mag}.
\label{eq:6}
\end{equation}

This discrepancy cannot be attributed to the contribution of companion
galaxies. In the case of PKS~1302$-$102, the companion at 2\arcsec\
lies inside the annular region studied (12~kpc to 25~kpc); if we
include its light, our estimated brightness for the host increases
0.2~mag. Part of the companion galaxy 5.5\arcsec\ from PKS~2135$-$147
lies in the annular region considered, but V\'eron-Cetty \& Woltjer
(1990) also subtracted its contribution from their measurements.  The
compact companion at 2\arcsec\ of PKS~2349$-$014 lies outside the
annular region considered.

The V\'eron-Cetty \& Woltjer magnitudes for the host that were
estimated by fitting
a spheroidal model are typically about one mag brighter than our 2-D model
magnitudes, i.e.,

\begin{equation} 
\langle{M_{V(i)}({\rm model})_{VCW} - M_{V(F606)}({\rm 2-D})}\rangle 
= -0.8 \pm 0.4~{\rm mag}\ .
\label{eq:7}
\end{equation}

Table~\ref{metric} lists the annulus measurements of the whole
sample.  For the range of redshifts
in our sample, the designated V\'eron-Cetty \& Woltjer annular 
region 12.5~kpc to 25~kpc ($H_0=50~ {\rm
km~s^{-1}Mpc^{-1} }$, $\Omega_0=0.0$) corresponds approximately to 6~kpc
to 12~kpc with our chosen cosmological parameters. 
In Table~\ref{metric} we list for each quasar the inner and
outer radii in arcsec, the apparent and absolute $F606$ aperture
magnitude, and the absolute $V$ magnitude in the annulus (see adopted
values for $F606-V$ values in Table~\ref{result}). As stressed by
V\'eron-Cetty and Woltjer (1990), these annular measurements can be
compared to future measurements obtained by other techniques.

\subsection{Dunlop et al.}
\label{subsecdunlop}

Dunlop et al.\ (1993) obtained deep ground-based near-infrared images
in the $K$ band for a sample of nearby $(z < 0.4)$ radio loud and
radio quiet quasars. They built a library of infrared PSFs by
observing many bright stars. The nuclear component was removed by
selecting from the library the PSF that produced the best match to the
quasar PSF. The stellar PSF was scaled to the same height as the
central peak of the quasar. Dunlop et al. suggest that their procedure
will cause the luminosities of the hosts to be overestimated, but in
practice the sign of the error could depend on whether there was a
positive or a negative fluctuation in the measured light in the
central peak. An aperture diameter of 12\arcsec\ was used by Dunlop
et al.  to measure the magnitudes of the hosts.  We have eight quasars
in common.

Table~\ref{dunlop} compares the $HST$ and the Dunlop et al.\
results. We list the $K$ magnitude they obtained for the host galaxies
using an aperture of diameter 12\arcsec, the $K$ absolute magnitude
for the host, using our comoslogical parameters, the color $(V-K)$ for
an elliptical galaxy obtained from Bruzual \& Charlot (1993), the
corresponding $V$ absolute magnitude expected, the 2-D model $V$ absolute
magnitude estimated from the $HST$-F606W images (see next-to-last
column of Table~\ref{bestmodel}),
 and the difference between the $V$ absolute
magnitude derived from both bands, $\Delta M_V$. The average
discrepancy is

\begin{equation} 
\langle~\vert~{\Delta M_V}~\vert~\rangle ~=~ \langle~\vert~{M_{V(K)}
- - M_{V(F606)}}~\vert~\rangle ~=~ 1.0 \pm 0.6~{\rm mag}.
\label{eq:8}
\end{equation}

For seven of the eight cases, our estimated magnitudes are brighter than
obtained by Dunlop et al.\ (1993).

\subsection{McLeod \& Rieke}
\label{subsecmcleod}

McLeod \& Rieke (1994a,b)  obtained ground-based images of luminous
quasars in the $H$-band. For most cases, they determined a 
one-dimensional profile for the quasar, subtracted a stellar PSF, and
then 
fit the resulting profile with an analytic galaxy model.  We have
fourteen quasars in common. To compare the results, we transform both
the $H$-band and the F606W magnitudes to the $V$ band. We used the
k-corrections and the relative sensitivities of the different bands
calculated by Fukugita et al.\ (1995) to convert our F606W measurements to
$V$ (see \S~\ref{secsize}).

To transform the $H$-band magnitudes to $V$, we assumed $(V-H) \sim3.0$ for
normal galaxies plus a k-correction given by McLeod and Rieke
(1995).  For objects not in their table, we interpolated
in redshift to obtain the expected $(V-H)$ for a normal galaxy.

Table~\ref{mcleod} lists the following information: column~1: quasar;
column~2: McLeod-Rieke $H$-band magnitude for the host galaxy;
column~3: color $(V-H)$ for normal galaxies including k-correction;
column 4: absolute $V$-magnitude based on the $H$-band measurements
and calculated assuming $H_0~=~100~ {\rm km~s^{-1}Mpc^{-1} }$, and
$\Omega_0$; column~5: absolute 2-D model $V$-magnitude
estimated from F606W images; and column~6: difference between absolute
$V$-magnitude estimated from $H$-band and F606W images. The average
discrepancy is

\begin{equation} 
\langle~\vert~{M_{V(H)} - M_{V(F606)}}~\vert~\rangle = 0.4 \pm
0.2~{\rm mag}.
\label{eq:9}
\end{equation} 

For 8 of the 14 cases, the McLeod \& Rieke luminosities are brighter
than the $HST$ luminosities. In some cases, the difference is clearly due to 
the
McLeod \& Rieke magnitudes also including the companion galaxies (see
PG~1012$+$008 in Figures \ref{figallqso} and \ref{figallqsosub}, 
and the discussion in
 \S~\ref{comments}).

\section{COMMENTS ON INDIVIDUAL CASES}
\label{comments}

In this section we will discuss the images of each of the quasars. 
Table~\ref{size} summarizes the morphological information obtained
from the $HST$ images and Tables~\ref{result}-\ref{bestmodel} give the
inferred luminosities of the host galaxies. For comparisons, the absolute
magnitudes  reported by other authors have been
converted to our cosmological parameters. Information about close
galaxy companions is summarized in Table~\ref{comps}.

Measurements of the surface brightness along the major axes of the
relatively bright elliptical host galaxies of PHL~909, PG~0923+201,
PKS~1004$+$130, HE~1029$-$140, PG~1116$+$215, 3C~273, and
PKS~2135$-$147, do not show evidence of discontinuity in the light
distribution. Thus all the apparently elliptical hosts discussed below
for which we could make detailed photometry satisfy the criterion of
having smooth light distributions.
The $HST$ images reveal spiral host galaxies with H~II
regions for three quasars, PG~0052+251, PG 1309$+$355, and
PG~1402+261.  It should be feasible to obtain spectra of the brightest
H~II regions. The spectra may reveal the composition of the material
which makes up the quasar hosts and perhaps provide further clues to
the quasar phenomenon.

\noindent  {\bf PG~0052$+$251:}\ 
The host is a beautiful spiral galaxy (see Bahcall et al 1996 for a
detailed discussion). The spiral host is evident even in our 200~s
exposure. 
The southern spiral arm extends the furtherest from the quasar (in the
direction of the companion).
There is good agreement between the absolute $V$
magnitude estimated from the McLeod and Rieke (1994b) $H$-band image
measurements and our 2-D model estimate based on the $HST$
images. Miller (1996, private communication) has recently obtained
spectra for some of the bright H~II regions identified (Bahcall et al.\
1996) in the $HST$ images. The H~II regions observed have the same
redshift as the quasar. The second and third quasar nuclei suggested
by Hutchings, Janson \& Neff (1989; the second nucleus was confirmed
by Dunlop et al.\ 1993), are seen on the $HST$ images to be,
respectively, a bright H~II~region in the spiral arm of the host
galaxy, and a foreground star in the Galaxy.

\noindent  {\bf PHL~909 (0054$+$144):}\ 
The quasar host is a normal elliptical E4 galaxy (see Bahcall et al
1996 for a more extensive discussion and a variety of images). This
radio quiet quasar does not occur in a spiral galaxy, as the
conventional view had suggested before the $HST$ observations. We did
not detect extended emission towards the western companion galaxy, as
suggested by Dunlop et al (1993). The elliptical host is apparent in
our 200~s image.

\noindent  {\bf NAB~0205$+$02:}\ 
The appearance of this quasar in the $HST$ images (see
Fig.~\ref{figallqso}) resembles, on visual inspection, that of a
bright star. Indeed, it is hard to distinguish the quasar from a star
by just looking at the images (see Figure~\ref{figallqso}).  After PSF
subtraction, we detect a small disk-like host galaxy
(see Figure~\ref{figallqsosub}), with size and inclination similar to
the companion 8.3\arcsec, position angle ${\rm PA=332^\circ}$. The
scale length of the host is only $\sim 1.2$~kpc and the absolute
visual magnitude is $-19.1$.  

Figure~\ref{bluevsred} compares the host image after subtraction of a
blue standard star, MMJ~6481 (see Figure~\ref{bluevsred}a), 
and after the subtraction of a red
standard star, F141 (see Figure~\ref{bluevsred}b).  The 
image of the host galaxy obtained after the subtraction of the PSF for
the blue
standard star is significantly clearer than the image obtained using
the red-star PSF.  However, there is very little difference
quantitatively, $< 0.1$ mag, between the host galaxy magnitudes we
determined from the two subtracted images.

Stockton, Ridgway, \& Kellogg (1996 in
preparation) recently obtained ground-based images in two line-free
continuum bands that show a definite host galaxy elongated roughly
NW-SE as well as some extended low-surface-brightness material to the
west.  In agreement with our results, they report that the host galaxy
has about the same luminosity and orientation as the companion in the
shorter wavelength image, but the host has somewhat redder colors than
the companion.  NAB~0205$+$02 was not resolved in the deep images
taken by Smith et al (1986).

Figure~\ref{nabobj} shows a fascinating object, first noticed by
Stockton and MacKenty (1987). The object, which is 12\arcsec\ to the
East of the quasar, is visible in their [O~III] image, but absent
from their continuum image.  It appears as a point-source with a
bright jet-like structure to the south and a much fainter and curved
extension to the north. The width of the jet and of the curved
extension is less than 0.3~kpc. The total apparent F606W magnitude of
the object is 22.6, and the extension of the jet-like structure and
curved tail are 0.7\arcsec\ (1.2~kpc) and 1.1\arcsec\ (1.9~kpc),
respectively. The offset of this object from the quasar position is
\hbox{$\Delta \alpha = 12.1$ \arcsec}\ and $\Delta \delta = 1.7$
\arcsec.

\noindent  {\bf 0316$-$346:}\ 
The morphology of the host environment is complex, probably
the result of gravitational interactions with a neighboring galaxy.
Figures~\ref{figallqso} and \ref{figallqsosub} show clearly what
appear to be tidal tails that extend about 20~kpc west of the
quasar. 
There are also bright diffuse clumps within $\sim 5$~kpc of
the quasar, which may contain H~II regions.  
We do not see unambiguous signs of  a
compact remnant close to the center of light of the quasar.  The
radial profile is reasonably described by an exponential disk from radius
$\sim$ 1\arcsec\ to 6\arcsec.  There is a relatively bright
peculiar spiral galaxy ($m_{606} = 20.7$~mag) projected 26.7\arcsec\
of the quasar, ${\rm PA=67^\circ}$.

\noindent  {\bf PG~0923$+$201:}\ 
The host of this radio quiet quasar is an E2 elliptical, a member of a
small group of galaxies. The two galaxies at 11.0\arcsec, 
${\rm PA=151^\circ}$, and 15.1\arcsec, ${\rm PA=162^\circ}$, have redshifts
similar to the quasar (Heckman et al 1984); the companion at 15\arcsec\
SE is not shown in Figures~1 and 2.  The $V$ magnitude estimated from McLeod
\& Rieke (1994b) $H$-band images is in good agreement (0.1~mag)
with the magnitude determined from our 2-D model.

\noindent  {\bf PG~0953$+$414:}\ 
Low surface brightness fuzz was detected in the early $HST$ images
(Bahcall et al. 1994, 1995a) and confirmed to be real by longer
(Cycle~5) exposures.  We are unable to establish the morphology of the
fuzz, which is faint and extended.  Hutchings et al (1989) suggested,
on the basis of ground-based images, that the host of the quasar has
spiral structure and is possibly interacting.  The HST images do not
provide convincing evidence for or against this conjecture.
Spectroscopy of the fuzz around the quasar obtained by Boroson et al
(1985), shows that the off-nuclear spectrum is dominated by a red
continuum with $H\alpha$ and possibly [S~II] emission; Mg~Ib
absorption might also be present.  The absolute $V$ magnitude derived
from McLeod and Rieke (1994b) $H$-band observations is 0.5~mag
brighter than our 2-D model estimate.

\noindent  {\bf PKS~1004$+$130:}\ 
The host is a bright elliptical, about as bright as the brightest
cluster galaxies.  
There seems to be some structure in the inner region of the host
($r < 2.5''$).
The absolute $V$ magnitude derived from McLeod and
Rieke (1994b) $H$-band observations is 0.8~mag fainter than our 2-D
model estimate.  Stockton and MacKenty (1987) presented narrow band
[O~III] images of this quasar, but did not find any significant
extended luminosity in the region between $\sim 7$ to 28~kpc (for $q_0
= 0.5$ and $H_0 = 100$ km s$^{-1}$Mpc$^{-1}$).  Stockton (1978)
obtained spectroscopic observations of the galaxies around the quasar
and found that two of them have redshifts similar to the quasar
redshift. One of the galaxies is 
separated by 33.4\arcsec\ (79.8~kpc, ${\rm
PA=233^\circ}$; not shown in Figures~1 and 2) from the quasar and has 
$m_{606}=20.0~{\rm mag}$; the other is not in
our $HST$ image field of view.

\noindent  {\bf PG~1012$+$008:}\ 
This quasar is ``caught in the act'' of merging.  The host of this
quasar is seen in Figures~\ref{figallqso} and \ref{figallqsosub} to be
merging with a bright galaxy. The distance between the two galactic
nuclei is 6.7~kpc (3.3\arcsec). There is another compact galaxy
6.8\arcsec\ (12.4~kpc) North of the quasar, probably taking part in
the interaction as well. Both interacting companions are also visible
in our 200~s exposure.

Figure~\ref{pg1012} shows an expanded view of the PG~1012$+$008 image.
Heckman et al.\ (1984)
obtained spectra for both galaxies and found the expected absorption
lines ($H\beta$, Mg~I, Fe~II and Na~I) at the same redshift as the
quasar.  The absolute $V$ magnitude based on the $H$-band measurement
by McLeod \& Rieke (1994b) is 0.9~mag brighter than our 2-D model
estimated value; we believe that part of this discrepancy is caused by
the light of the merging galaxy being included in their model fit.

\noindent  {\bf HE~1029$-$140:}\ 
This radio quiet quasar has a bright elliptical host galaxy. Some
faint structures resembling shells are seen at $\sim 11\arcsec$ and
19\arcsec\ from the quasar. There is a compact galaxy 4.1\arcsec\
North from the quasar; spectroscopy is required to determine if it is
associated with the quasar. Wisotzki et al.\ (1991) obtained $R$-band
images and, based on spheroidal model fits, reported the host to be a
giant elliptical with $R = 15.2$~mag and $M_R = -22.0$~mag. Assuming
$(V-R) \sim 0.8$ for elliptical galaxies at $z
\sim 0.1$ (Fukugita et al. 1995), their absolute $V$ magnitude for the host
is $-21.2$~mag. We obtained $M_V =-20.8$ by fitting a GdV
model to the azimuthally averaged profile of the host. 

Wisotzki (1994) measured redshifts of galaxies in the field of
HE~1029$-$140 and found that four galaxies have redshifts similar to
the quasar. Wisotzki did not detect the close compact object at
4\arcsec\ projected separation.

 The closest galaxy with a redshift known to be similar to the quasar
lies at 134~kpc (=109.4\arcsec) ${\rm PA=26^\circ}$ and was detected
in WF2 (not shown in Figures~\ref{figallqso} and \ref{figallqsosub}).
The $HST$ images show that this Wisotzki companion galaxy
has $M_V \sim -19.1$ ($m_{606} =17.7$) and 
is highly disturbed, probably an advanced stage of merger between two
disk galaxies.  Two possibly galactic nuclei, separated by 1.4\arcsec,
are visible, as well as H~II~regions. 

There is a $m_{606}=17.9$~mag galaxy projected at 22.9\arcsec, ${\rm
PA=23^\circ}$, of the quasar, which Wisotzki showed is a background
galaxy with $z=0.162$. Wisotzki classified this galaxy as an
elliptical, but the $HST$ images show that it is a spiral galaxy (type
$\sim$Sab).

\noindent  {\bf PG~1116$+$215:}\ 
The host of this radio quiet quasar is probably an elliptical
galaxy. The bright central region is surrounded by a faint smooth
fuzz. Some arc-like structures are seen in the center, but are almost
certainly PSF artifacts.  
The one-dimensional profile is reasonably
described by a de Vaucouleurs model.
The $V$ magnitude estimated
from McLeod \& Rieke (1994b) $H$-band images is in good agreement
with the magnitude determined from our 2-D model.

 McLeod and Rieke (1995) applied the techniques they developed to
study their $H$-band ground-based images, to reanalyse the $HST$
Archive images of PG~1116$+$215. They subtracted just enough of the
PSF of the red standard star, F141 ($B - V = 1.11$),
from the azimuthally-averaged quasar 1-D profile to make the
residual profile almost turn over in the center. Because the $HST$
images were saturated in the center, McLeod \& Rieke normalized the
light profile just outside the saturation region. We
note, however, that the region inside $r\sim 2\arcsec$ is heavily
contaminated by PSF artifacts when the PSF of the red star F141 is
used for the subtraction. If F141 is used for the PSF subtraction, 
there are artifical radial spikes - which are not
visible in their published $HST$ image of this quasar. 
McLeod and Rieke state that the host of PG~1116$+$215
``looks nearly identical" to a field galaxy detected in WF4. An $HST$
image of this field galaxy was published in Bahcall et al.  (1995a, Figure
5c). The field galaxy  is rich in morphological details:
it is a barred spiral with internal and external rings (~RSBb).  If
it is at the same distance as the quasar, its absolute $V$ magnitude
is $-20.9$. Although the host and the field galaxy have apparently
similar luminosities, they are morphologically quite different.

\noindent  {\bf PG~1202$+$281:}\ 
The host of this radio quiet quasar, also known as GQ COM,
 is a small elliptical E1 galaxy.  
Stockton and MacKenty (1987) showed that the compact galaxy
 $\sim 5\arcsec$ from the quasar, ${\rm PA=71^\circ}$, is at the same
 redshift as the quasar. Boroson et al.\ (1985) reported that the off-nuclear
 spectra are dominated by a red continuum, with [O~III] lines and
 possibly $H\alpha$ in emission, and a possible Mg~Ib absorption
 feature. The $V$ magnitude estimated from the McLeod \& Rieke (1994b)
 $H$-band images is in good agreement with the magnitude determined by
 our 2-D model.

\noindent  {\bf 3C~273  (PG~1226$+$023):}\ 
The host galaxy is an elliptical.  The $V$ magnitude estimated from
the McLeod \& Rieke (1994b) $H$-band images is in good agreement with
the magnitude determined by our 2-D model.  Using deep ground-based
CCD images, Tyson et al.\ (1982) obtained $M_V \sim -22.5$ for the
host, which is in good agreement with our best 2-D model magnitude of
$M_V=-22.1$.
The host is somewhat brighter than the brightest galaxy in a rich cluster.
Stockton (1980) measured redshifts for galaxies in the 3C~273
field and found that four of them have redshifts similar to the
quasar, in agreement with the suggestion of Bahcall \& Bahcall
(1970). One of those galaxies was detected (a spiral galaxy) in WF4:
it lies at 75\arcsec\ East of the quasar ($\sim133$~kpc) and its
redshift is $z=0.1577$. Wyckoff et al.\ (1981) obtained $R = 16.3$ for
the host, which is equivalent to $M_V \sim -21.3$~mag. The inner part
of the jet is barely visible in the PSF subtracted image in
Figure~\ref{figallqsosub}.  $HST$ and Merlin observations of
the 3C~273 jet are reported in Bahcall et al.\ (1995).

\noindent  {\bf PKS~1302$-$102:}\ 
The $HST$ images show there are two small compact galaxies at
1\arcsec\ and 2\arcsec\ from the quasar, which are expected to spiral
into the quasar in a time short compared to the Hubble time (Bahcall et al.\
1995a).  The presence of these very
close companions can be seen most clearly in the expanded image,
Figure~8, of Bahcall et al.\ (1995a).  Hutchings and Neff (1992)
performed optical imaging with 0.5\arcsec\ resolution and reported
structures at the positions of those galaxies; they suggest the host
is a mildly disturbed elliptical galaxy.  
The PSF-subtracted residual image (see Figure~\ref{figallqsosub}) 
appears to the eye to be similar to an elliptical galaxy,
 but an exponential
disk fits the data slightly better. The absolute $V$ magnitude estimate for the
host galaxy is 0.4~mag fainter than the value determined from the McLeod
\& Rieke (1994b) $H$-band observations, and is 0.9~mag fainter than
the value estimated from $i$-band images (V\'eron-Cetty \& Woltjer
1990).  Wyckoff et al.\ (1981) obtained $R = 19.0$ for the host; using
the Fukugita et al.\ (1995) transformations of galaxy colors, this
corresponds to $M_V
\sim -20.0$.

\noindent  {\bf PG~1307$+$085:}\ 
The host appears to be a small early-type galaxy. 
The absolute $V$ magnitude estimated from our 2-D model of the
$HST$ image is 0.2~mag fainter than the value derived from the McLeod
\& Rieke $H$-band images.

\noindent  {\bf PG~1309$+$355:}\ 
Figure~\ref{pg1309} shows an expanded view of the image for
PG~1309$+$355 with lower contrast than shown in
Figures~\ref{figallqso} and \ref{figallqsosub}. The  details near the
center of PG~1309$+$355 are shown more clearly in Figure~\ref{pg1309},
after a best-fit stellar PSF has been subtracted.
This quasar has a bright early-type host galaxy, probably an Sab.
Tightly wrapped spiral arms are clearly seen in the inner region
surrounding the center of the quasar.
Overall, the GdV model describes the radial
profile better than the exponential disk model.  
The absolute $V$ magnitude estimated from
our 2-D model of the $HST$ image is 0.1~mag brighter than the value
derived from the McLeod \& Rieke (1994b) $H$-band images.

\noindent  {\bf PG~1402$+$261:}\ 
Figure~\ref{pg1402} shows an expanded view of the image of this quasar
at two different contrast levels.  The $HST$ images show that the host
is a beautiful spiral.  After the PSF subtraction a bar and a possible
inner ring are visible. The morphological type is approximately
SBb(r?). H~II regions are also visible along one of the spiral arms;
they are less prominent than the H~II regions seen in the spiral host
of PG~0052$+$251.  The relative positions and magnitudes of the
brightest H~II regions are marked for identification in the upper
panel in Figure~\ref{pg1402}. Table~\ref{hiitab} lists for each H~II
region the aperture magnitude, the distance to the quasar nucleus, and
the offset in right ascension and declination between the H~II region
and the quasar nucleus. The magnitudes were measured using apertures
of $0.3\arcsec$.

PG~1402$+$261 is a relatively isolated quasar.  Stockton and MacKenty
(1987) did not find any significant extended [O~III] emission around
the quasar.  McLeod and Rieke (1994b) obtained $H$-band images; the
absolute $V$ magnitude based on their measurement is 0.5~mag brighter 
than the value estimated with our 2-D model of the $HST$ images.

\noindent  {\bf PG~1444$+$407:}\ 
The host galaxy of this quasar is smooth and is elongated in the NS
direction.  After the PSF subtraction, the nuclear region is seen as
an extended structure running NE-SW, as shown in
Figure~\ref{figallqsosub}. A bar may be present.  The exponential disk
model describes the radial light distribution a little better than
does the GdV model.  The overall appearance is most similar to an E2
galaxy.  Hutchings and Neff (1992) originally suggested the
possibility of a bar in this host, which they proposed might be in an
advanced stage of the merger of galaxies of very different masses.
The absolute $V$ magnitude estimated from our 2-D model of the $HST$
image is 0.5~mag fainter than the value derived from the McLeod \&
Rieke $H$-band images.

\noindent  {\bf 3C~323.1 (PG~1545$+$210):}\ 
A low surface-brightness elliptical host galaxy appears in the $HST$ images, as
well a neighboring companion at a projected distance of $\sim 7$~kpc,
(2.7\arcsec, ${\rm PA=292^\circ}$),
(see Figures~\ref{figallqso} and \ref{figallqsosub}).  The $V$
magnitude of the host galaxy determined from the $H$-band images by
McLeod and Rieke (1994b) is 0.4~mag brighter than the value derived
from our 2-D model of the $HST$ images. More recently Neugebauer
et al.\ (1995) obtained $H$ and $K$-band images of 3C~323.1; their
estimated model fit $H$-band magnitude for the host agrees with McLeod
\& Rieke (1994b) results. Neugebauer et al.\ measurements suggest that
if the host galaxy is a normal elliptical, its expected apparent $V$
magnitude is $\sim 17.5$, and if it is an spiral galaxy, $V \sim
16.9$. These values are brighter than the $V$-band apparent magnitudes
expected using a color transformation without reddening of 
our $HST$ images (see Table~\ref{model}), but are
consistent with the host galaxy being a reddened elliptical, as
pointed out by Neugebauer et al.\ (1995). 
The GdV one-dimensional model fits better the surface
brightness radial profile of the quasar host in our $HST$ image than
does the exponential disk model.

 Stockton (1982) took spectra of the companion galaxy at 2.7\arcsec\ 
from the quasar, and apparently detected [O~III] at the same
redshift as the quasar.  Complex, asymmetric extended [O~III] emission
is seen surrounding the quasar in [O~III] images obtained later by
Stockton and MacKenty (1987).  However, there is no excess of emission
at the companion galaxy position in the [O~III] images. Stockton and
MacKenty conclude that the previously reported detection of [O~III]
emission in the spectrum of the companion galaxy was fortuitous, since
the emission is due to the general distribution of ionized gas
surrounding the quasar. 
Neugebauer et al.\ found that the $H$ and $K$
model fit magnitudes obtained for the companion galaxy, when combined
with the $HST$ measurements by Bahcall et al.\ (1995a), gives a color
$V-H$ consistent with the companion galaxy being a faint elliptical at
the redshift of the quasar.

\noindent  {\bf PKS~2135$-$147:}\ 
The quasar host is an elliptical galaxy; the envelope is not
featureless like PHL~909, but contains some faint clumps.
The image suggests the presence of a jet at 
${\rm PA \sim 26^\circ}$, with visual extension of $\sim 2.4$\arcsec\ 
 and width $\sim 0.5$\arcsec.
The surface brightness of the jet-like feature is 
 $23.4~{\rm mag~arcsec^{-2}}$ (F606W), at 1.7\arcsec\  from the quasar
center of light.
 Many galaxies are seen in the
field; the closest companions to the quasar are at 1.9\arcsec, ${\rm
PA=128^\circ}$, and 5.5\arcsec, ${\rm PA=119^\circ}$, seen in
Figures~\ref{figallqso} and \ref{figallqsosub}.  Stockton (1978, 1982)
obtained spectra of the field galaxies and found that four are at the
same distance as the quasar, including the closest companions at
projected separations of 1.9\arcsec\ and 5.5\arcsec. Somewhat later,
Stockton \& MacKenty (1987) obtained [O~III] images. They noted that
the companion 2\arcsec\ southeast did not appear enhanced in the
[O~III] images, indicating that the emission lines seen in the
combined spectrum are due to extended emission around the quasar that
is not confined to the companion.  Hickson and Hutchings (1987)
reported Mgb in the spectrum of the secondary nucleus (the galaxy at
2\arcsec\ SE) corresponding to a galaxy at the redshift of the quasar.

PKS~2135$-$147 has been included in many imaging programs.  For
example, Dunlop
et al.\ (1993) measured $M_K({\rm host}) =-24.2$, which gives $M_V \sim
- -21.0$.  Our 2-D model analysis of $HST$ images gives $M_V \sim -21.1$.
V\'eron-Cetty \& Woltjer (1990) obtained $i$-band images and measured
aperture photometry in an annulus with radii 12.5~kpc and 25~kpc. They
estimated $M_V= -19.9$. We measured the absolute magnitude of the host
in our $HST$ images within the same annulus and obtained $M_V \sim
- -19.4$.  Smith et al.\ (1986) measured the absolute blue magnitude to
be $M_B \sim-21.5$, which converts to $M_V \sim-19.9$ (if $(B-V)\sim
1.6$ for an elliptical galaxy at $z=0.2$).  Wyckoff et al.\ (1981)
obtained $R=17.8$, which corresponds to $M_V \sim -19.3$.

\noindent  {\bf PKS~2349$-$014:}\ 
The quasar host galaxy is undergoing gravitational interaction, which
is evidenced by apparently tidal arms and possibly also by a huge
($\sim 50$~kpc) diffuse off-center nebulosity. A compact companion
galaxy at 2\arcsec\ E is detected in the $HST$ images. Bahcall et al.\
(1995b, 1995c) provide an extensive discussion of this object and
include a number of different images (the tidal wisps and extended
nebulosity are seen most clearly in their Figures~\ref{figallqso} and 
\ref{figallqsosub}; the close
companion galaxy is featured in their Figure~3).  Dunlop et al.\ (1993)
suggests that the quasar is interacting with the three closest
galaxies SE of the quasar. The $HST$ images do not show evidence for
this interaction; the three galaxies do not seem to be morphologically
disturbed and no obvious link between them and the quasar is seen.
Dunlop et al.\ measured $M_K(host) =-24.7$, which gives $M_V \sim
- -21.5$. Our 2-D model fits to the $HST$ image of the host gives $M_V
\sim -22.0$, similar to the light from a brightest cluster galaxy.

The radial profile of the interacting system PKS~2349$-$014 is well
described by a de Vaucouleurs model.  As pointed out by Toomre (1995)
and Bahcall
et al.\ (1995b), there are some morphological resemblances between
PKS~2349$-$014 and NGC~3921. Schweizer (1996) has carefully and
extensively studied NGC~3921, suggesting that it is the result of a
merger between two disk galaxies and is now a protoelliptical. In this
case also the mean light distribution of the system is well described
by a $r^{1/4}$\ law.

\section{SUMMARY AND DISCUSSION}
\label{discuss}

The images shown in Figures~1 and 2 establish the two main conclusions
of this paper: 1) The most luminous nearby quasars exist in a variety
of environments; 2) $HST$ observations provide unique information
about the circumstances of the quasar phenomenon. Many of the most
important results obtained in this paper are visible on the
unprocessed images shown in Figure~\ref{figallqso}, which can be
compared with the images with a stellar PSF subtracted shown in
Figure~\ref{figallqsosub}. The subtraction of a stellar PSF is
important in about half of the cases.

Our results are based upon a representative sample of 20 of the most
luminous known ($M_V < -22.9$) and nearby ($z < 0.30$) quasars.
The characteristics of
the sample are summarized in Table~\ref{journal} and
Section~\ref{subsample}.

In separate subsections, 
we summarize and discuss below our results on  host morphologies, 
host luminosities,    comparisons with  our  earliest
analyses,   close companions,   ground-based studies, and 
future work.

\subsection{Host Morphologies}
\label{subhostmorpho}

Figures~\ref{figallqso} and \ref{figallqsosub} show three hosts that
apparently are normal spirals with H~II regions (PG~0052$+$251,
PG~1309$+$355, and PG~1402$+$261), seven ellipticals (PHL~909,
PG~0923$+$201, PKS~1004$+$130, HE~1029$-$140, PG~1116$+$215, 3C~273,
and PKS~2135$-$147), as well as three obvious cases of current
gravitational interaction (0316$-$346, PG~1012$+$008, and
PKS~2349$-$014).  There are also five other hosts that appear to be
elliptical galaxies and are listed as En(?) in Table~\ref{size}.  The
hosts for two of the quasars (NAB~0205$+$02 and PG~0953$+$414) are
faint and difficult to classify. 

The host galaxies are centered on the
quasars to the accuracy of our measurements, $\pm 0.4$~kpc (see
Section~\ref{secsize} and Equation 6).

Seven of the 14 radio quiet quasars
in our sample occur in hosts that are classified as elliptical
galaxies in Table~\ref{size}.  Two particularly beautiful examples of
elliptical hosts for radio quiet quasars are shown in Figures 1 and 2
for PHL~909 and HE~1029$-$140.  We have presented a more extensive
discussion of the host of PHL~909 in Bahcall et al.  (1996).  Five of
the six radio loud quasars in our sample appear to lie in elliptical
galaxies. The sixth radio loud quasar, PKS~2349$-$014, is in a complex
interacting system containing a close companion, apparent tidal tails,
and a large off-center nebulosity.

The fact that about half of the radio quiet quasars in our sample have
elliptical hosts contradicts the conventional wisdom that radio quiet
quasars occur in spiral galaxies. However, we confirm the expectation
that most radio loud quasars are in elliptical galaxies.

Three of the  quasars in our sample, 0316$-$346,
PG~1012+008, and PKS~2349$-$014, 
 have been ``caught in the act'', i.  e., the $HST$ images of these
quasars show dramatic evidence of currently intense gravitational
interactions.  Two of the three quasars caught in the act are radio
quiet.  In all three cases, the unprocessed images (see
Figure~\ref{figallqso}) are sufficient to reveal extended curved
features that look like the tidal arms generated in numerical
simulations of interacting galaxies. 
For  0316$-$346 and
PKS~2349$-$014, there is no clear
evidence for a normal host galaxy centered on the quasar.
PG~1012+008 appears to be an example of a merger currently
going on between two comparable galaxies.  For a more extensive
analysis of PKS~2349$-$014, the reader is referred to Bahcall et
al. (1995b,c), in which the close companion, the tidal arms, the very
extended 
off-center nebulosity, and the possible host galaxy are all discussed
in some detail.

The HST images provide the best available data base to search for
optical jets in nearby luminous quasars.  Information on the existence
or non-existence of small scale optical jets can constrain theories of
the origin of radio and optical jets.
We examined carefully 
the images of all 20 of the quasars for the existence of
optical jets.  
With the exception of the well-known jet in 3C~273, the only other
quasar for which we have found evidence for linear features is 
PKS~2135$-$147.  Sensitive radio searches should be undertaken to test
whether PKS~2135$-$147 has a radio jet.
For the other quasars, 
we can rule out the existence of a narrow optical feature
with a surface brightness in excess of 24.5 mag~${\rm sec^{-2}}$
extending more than 3 kpc beyond an inner region beginning at about 6
kpc.  

\subsection{Host Luminosities}
\label{subhostlum}

We list in Table~\ref{summary} our best estimates for the
magnitudes of the host galaxies in our sample; these magnitudes were
obtained (see \S~\ref{subtwo}) by fitting a two-dimensional analytic
galaxy model to the data.  We also list the effective radius or
exponential scale length, and give the morphology of the host based on
visual inspection of the images (see Table~\ref{size} and
\S~\ref{subunproc}).  The different measurements of the brightnesses
of the host environments are summarized and compared in
Tables~\ref{result} and \ref{model} and in Section~\ref{secsize}.
The average best-fit 2-D model magnitudes for the hosts of the 14
radio quiet quasars is

\begin{equation}
\langle M_V \rangle _{\rm model,radio~ quiet} = -20.6 \pm 0.6~{\rm mag} .
\label{eq:10}
\end{equation}
The hosts of the 6 radio loud quasars are slightly brighter, 

\begin{equation}
\langle M_V \rangle _{\rm model,radio~ loud} = -21.6 \pm 0.6~{\rm mag} .
\label{eq:11}
\end{equation}

The fact that in our sample the radio loud quasars are, on the
average, about a magnitude brighter than the radio quiet quasars
cannot be explained by a selection effect
 resulting from the fact that the
radio loud quasars have a slightly larger average redshift. In fact,
the 3 radio loud quasars with $z \le 0.20$ have $\langle M_V \rangle =
- -21.8$~mag and the 3 radio loud quasars with $0.20 < z < 0.30$
have $\langle M_V \rangle = -21.4$~mag.

Figure~\ref{mvmv} shows the  2-D model  absolute 
visual magnitudes (from Table~\ref{summary})
of the host galaxies (and other nebular material) 
versus the absolute visual magnitudes of the
quasars. 
For two of the quasars, PHL~909 and 3C~323.1,
the symbols overlap at $M_V({\rm host}) = -21.2$ and
$M_V({\rm QSO}) = -22.9$.

In order to be detectable, the host must have a
luminosity that is not too small when compared with the luminosity of
the quasar.  The minimum detectable host brightness depends strongly
upon the assumed morphology of the host galaxy.
We have shown by a series of numerical experiments, described
in Table~3 of Paper~II, that host galaxies are, on the average, visible
on our images down to about 4.2~mag fainter than the quasar
luminosity.  

Galaxies that are smooth ellipticals are the most difficult to detect
(see rows
5d and 5e of Table~3 of Paper~II). 
For the eight quasars discussed in Paper~II,
smooth elliptical hosts are, on
average,  visible on
our images down to  $3.5 \pm 0.5$ mag fainter than the quasar.
The limiting brightnesses 
were determined by visually inspecting simulated galaxies
placed in the actual HST quasar observations and are therefore
somewhat subjective.

The diagonal line in Figure~\ref{mvmv} represents the detection limit
for smooth ellipticals in an idealized sample in which the limiting
host magnitude is determined entirely by the quasar luminosity.  
In calculating the limiting absolute visual magnitudes for the hosts, 
we have included an average k-correction (see Fukugita et al. 1995) 
for elipticals at $z = 0.2$,
as well as the average magnitude difference, $3.1$ mag, between the
quasar and the faintest detectable host.
Thus

\begin{equation}
M_{\rm limiting~host} = M_{\rm QSO} + \Delta.
\label{linearrelation}
\end{equation}
The form of Equation~(\ref{linearrelation}) reflects the fact that the
limiting host luminosity must increase
linearly with the luminosity of the quasar, since by assumption
the noise introduced by   the quasar signal determines how
faint a host can be detected.
Smooth ellipticals  fainter than $\pm 0.5$ mag
of the diagonal line in Figure~\ref{mvmv} would 
not have been
expected to be detected if this idealization of the problem is correct.
For the three intrinsically faintest quasars in our sample, the noise
introduced by the quasar, the host galaxy, and the background light
are all similar.  In practice, for our sample 
one might expect some flattening of the
detection limit at the lowest quasar luminosities if photon noise is
more important than systematic uncertainties in the subtraction of the
quasar light. 

McLeod and Rieke (1995) have suggested
that there is a linear relation between the quasar absolute magnitude
and the minimum host galaxy absolute magnitude. They interpret this
linear relation, shown in their Figure~2, as indicating that a more
luminous host galaxy is required to fuel a more luminous quasar.  The
linear relation that they find between $M({\rm host})$ and $M({\rm
QSO})$ is essentially identical to our minimum detection limit for
smooth ellipticals that is shown
in Figure~\ref{mvmv} (for $V - H =  3.0$). 
As pointed out  by McLeod(1996), the relation described by McLeod and
Rieke cannot be an artifcat produced by  detection limits if all of their
detections are real detections.  An artifical correlation would be
introduced only if true non-detections were interpreted as marginal
detections.

There is not convincing 
 evidence in Figure~\ref{mvmv} for a significant dependence of host
luminosity upon the luminosity of the quasar.  The apparent
correlation that is suggested to the eye is due in large part  
to the fact that the single 
most luminous quasar in our
sample, 3C~273, has the most luminous host.

There is a hint in Figure~\ref{mvmv} that the average luminosity of
elliptical hosts is somewhat higher than for spiral hosts.  Most of
this difference, however, is due to the fact that, even for the same
objects (cf. Table~\ref{model}),  the de Vaucouleurs
fitting formula yields estimated luminosities that are $0.6$ mag
brighter than the disk fitting function.  The de Vaucouleurs formula
introduces a luminosity peak in the unmeasured region that is not
present in the disk formula.

Our results are inconsistent with the hosts having a Schechter
luminosity function.  The average absolute magnitude for a field
galaxy is about 1.8 mag fainter than $M_V(L^*) = -20.5$, (for an
assumed minimum luminosity of $M_V(L^*) = -17$; see, e. g.,
Efstathiou, Ellis, and Peterson 1988 for a discussion of the field
galaxy luminosity function).  In our sample (see Table~\ref{summary}
and Figure~\ref{mvmv}) the average host is $M_V(L^*) = -20.9$.
Moreover, about half by number of the field galaxies would be
expected, in a volume limited sample, to lie within a magnitude of the
lower limit cutoff of the Schechter luminosity function (which may be
fainter than $M_V(L^*) = -17$).  Thus, if the host galaxies were
distributed with a Schechter luminosity function, we would have
expected that about half of the hosts in our sample would be fainter
than $M_V(L^*) = -18$ and therefore undetectable on the HST
images. This is clearly not the case.

By comparing the 2-D model magnitudes of Table~\ref{summary}
 and Figure~\ref{mvmv} with
the results expected from a Schechter luminosity function, we conclude
that, on average, the host galaxies of the luminous quasars in our
sample are about $2.2$ magnitudes more luminous than typical field
galaxies.

\subsection{Previous Analyses}
\label{subprevious}

The conclusions presented in this summary paper are different in
emphasis from the
conclusions in our first two studies (see Papers~I and~II).
In our earlier work, we discussed images of eight quasars (all are included
in the sample in the current work), and reported the definitive
detection of  three host galaxies. We also presented limits on the 
brightnesses of the hosts for the other five quasars. The images presented in
this paper show that more than half
of the entire sample of 20~quasars
has obvious hosts, and there is solid evidence
that most, if not all, of the remaining quasars also have host galaxies.

The initial caution that we expressed regarding the
detection of host galaxies was due to a combination of the unlucky
observing sequence and our conservatism about interpreting the complex
$HST$ images.  The unprocessed $HST$ data (see Figure~\ref{figallqso})
are sufficient to show, even to the untrained eye, that at least nine
of the 20 quasars in our sample have obvious hosts or diffuse
environments.  These obvious examples include the three spiral hosts
(PG~0052+251, 1309+355, and PG~1402+261), the three quasars ``caught
in the act'' (0316$-$346, PG~1012+008, and PKS~2349$-$014), and three
prominent ellipticals (PHL~909, HE~1029$-$140, and 3C~273).  None of
these quasars were among the first four objects (PG~0953+414,
PG~1116+215, PG~1202+281, and PG~1307+085) observed (Paper~I), and
only one (3C~273, whose host we described in paper~II together with
the original observations) belonged to the sample of eight quasars
(Paper~II). 
With a  probability of $9/20$ per observation of observing an obvious
host, 
it was simply bad luck (about a 5\% chance) that only one 
of the initial eight quasars
studied in Paper~I and Paper~II  had an obvious host.

Given  the repaired, but still complex and temporally variable PSF
of the $HST$, we presented our results (see Table~3 of Paper~II) for
the non-detections as a morphology-dependent limit on the brightness
of the host galaxy.  Spiral galaxies, with their azimuthal variation
in brightness and regions of high surface brightness, could be seen to
considerably fainter total brightnesses (more than a magnitude) than
large, extended ellipticals with their smooth, regular profiles.  As
we have gained experience with the data, we have become more confident
of our ability to judge the reality of low surface brightness
features.  Most importantly, during $HST$ Observation Cycle~5, we
obtained additional images of PG~0953+414, a quasar analyzed in
Paper~I which showed very faint, extended nebulosity that was not
centered on the quasar.  The Cycle~5 observations were obtained at a
different roll angle than those described in Paper~I; the new
observations confirmed the reality of the diffuse features (the
nebulosity remained fixed in the sky when the telescope was rotated).
The observations of PG~0953+414 suggested that some of the faint
features were real that we had initially worried could be PSF
features.

In retrospect, it appears that the vast majority of our initial
observations consisted of quasars whose hosts had smooth, regular
profiles.  Comparing our adopted brightnesses of the hosts
(Table~\ref{summary}) with the brightness limits set in 
 our earlier work by simulations
(Table~3 in Paper~II), we find that the appropriate brightnesses
limits (those for smooth ellipticals like 5d and 5e of Figure~5 of
Paper~II) were reasonably accurate.
The  detected brightnesses reported in Table~\ref{summary} of this paper
range from considerably fainter
than the Paper~II limit ({\it e.g.,} PG~0953+414) to
slightly brighter than the limit ({\it e.g.,} 3C~323.1).  Since the
morphology of the hosts of the initial quasars was biased towards one
type of galaxy (the least favorable type as far as detectability), it
was not accurate, as we did, to quote a detection
limit that was the average of all the galaxy types.

\subsection{Close Companions}
\label{subclosecomp}

The $HST$ images frequently reveal companion galaxies that are
projected very close to the quasar, in some cases as close as $1''$ or
$2''$. Table~\ref{comps} shows 20 galaxy companions that are projected
closer than $25$ kpc to the center-of-light of a quasar and brighter
than $M_({\rm F606W}) = -16.4$.  Altogether, 13 of the 20 quasars in
our sample have close companions that satisfy the requirements for
inclusion in Table~\ref{comps}. The amplitude for the quasar-galaxy
correlation function determined from our data is $3.8 \pm 0.8$ times
larger than the galaxy-galaxy correlation function (Fisher et
al. 1996).

\subsection{Ground-based Studies}
\label{subground}

Our results for individual objects are compared in
Section~\ref{groundsection} with the results from previous
ground-based observations. 
In general, the agreement with ground-based observations is
satisfactory, but not as precise as we would have hoped. The most
straightforward comparison is with the results of V\'eron-Cetty \&
Woltjer (1990) in annular regions well separated from the quasar. Even
in this case, our $HST$ magnitudes are, on average, 0.4~mag fainter
than the V\'eron-Cetty \& Woltjer values. Our 2-D model estimates for
the total luminosities are, on average, 0.8 magnitude fainter than
their 1-D model fits. The average discrepancy between our results and
Dunlop et al.\ (1993) is $1.0 \pm 0.6$~mag (our results are generally
brighter than Dunlop et al.), and the average discrepancy between our
results and McLeod \& Rieke is $0.4 \pm 0.2$~mag.

\subsection{Future Work}
\label{subfuture}

Some pre-HST theoretical analyses (e. g.,by  Falle, Perry, and Dyson 1981,
Weymann et al. 1982, 
Begelman 1985, and Chang, Schiano, and Wolfe 1987) have suggested that
luminous quasars may have dramatic effects on their environments via
the radiation 
or  hot winds that the quasar emits.  
The continuum images shown in this
paper do not provide  obvious 
evidence for the effects of the quasar on the
host environment. In fact, the three spiral host galaxies and several
of the host ellipticals appear remarkably normal.
Broad band colors and spectroscopic observations are required in order
to determine more sensitively whether the host galaxy is really
oblivious to the presence of the luminous quasar in its center. 
The theoretical modeling can now be made  more specific and 
compared with the results of the HST observations for individual host
galaxies. 
These studies will be important in constraining  the time scale of the
quasar phenomenon and the isotropy of the quasar emissions.  If a
quasar shines brightly for only a short period of time or if the
emission is highly anisotropic, then the lack of a dramatic effect of the
bright AGN on the surrounding medium may be more easily
understood.

One of the key results that is apparent in Figures~\ref{figallqso} and
\ref{figallqsosub} is the detailed evidence for gravitational
interactions among the systems ``caught in the act'' 
 (0316$-$346, PG~1012+008, and PKS~2349$-$014).  Dynamical modeling of
these systems could provide 
insights into the processes involved in the 
formation and fueling of quasars.

$HST$ images  provide detailed
quantitative information about the environments in which the quasar
phenomenon occurs.  We hope to increase our sample in the future and
to obtain color information about the objects discussed
here.  It should be feasible to obtain spectra of the brightest H~II
regions in the spiral hosts of PG~0052+251, PG~1309+355, and
PG~1402+261.  Detailed  comparisons between the $HST$ and
ground-based images will be very informative.

\acknowledgments
We are grateful to P. Crane, M. Fall, O. Gnedin, K. Kellermann,
K. McLeod, J. Miller, G. Neugebauer, 
M. Rees, G. Rieke, and M. Strauss
for valuable discussions, comments, and
suggestions.  
J. Krist and C. Burrows constructed the four stellar PSFs used in this
work; we are grateful to them for their skill and generosity in
providing these data.
We would like to thank Digital Equipment Corporation for
providing the DEC4000 AXP Model 610 system used for the
computationally intensive parts of this project.  This work was
supported in part by NASA contract NAG5-1618, NAG5-3259, NASA grant number
NAGW-4452 and grant number GO-5343 from the Space Telescope Science
Institute, which is operated by the Association of Universities for
Research in Astronomy, Incorporated, under NASA contract NAS5-26555.
We have used the NASA/IPAC Extragalactic Database (NED),
operated by the Jet Propulsion Laboratory, Caltech, under contract
with NASA, and NASA's Astrophysics Data System Abstract Service (ADS).
\vfil\eject

\newpage

\centerline{\bf FIGURE CAPTIONS}

\begin{figure}[h]
\caption[]{\baselineskip=16pt A $23\arcsec \times 23\arcsec$\ WF image of each 
one of the twenty
luminous nearby quasars in our sample. A blue field star, MMJ~6490, is
also shown for comparison (first panel). These images were obtained
using the {\it HST}\ WF3 and the F606W filter. The exposure times are
1400~s or 1100~sec (see Table~\ref{journal}). Cosmic ray subtraction
and pipeline STScI flatfielding are the only processing performed on
the $HST$ images shown here.\protect\label{figallqso}}
%\end{figure}
%\nopagebreak
%\begin{figure}[h]
\caption[]{\baselineskip=16pt Same set of quasars and 
comparison blue star as shown in
Figure~\ref{figallqso}, but in this case the best-fit PSF for
a standard blue star has been
subtracted.\protect\label{figallqsosub}}
\caption[]{\baselineskip=16pt The difference between using 
a PSF for a blue star and a PSF for a red star.  
The image in panel a) was obtained by 
subtracting a PSF of MMJ~6481 ($B - V = -0.07$) and the image in panel
b) was obtained by subtracting a PSF of F141 ($B - V = 1.11$).
The measured host galaxy magnitudes are the same within $0.1$ mag for
the two subtractions shown. \protect\label{bluevsred}}
%\end{figure}
%\nopagebreak
%\begin{figure}[h]

%\end{figure}
%\nopagebreak
%\begin{figure}[h]
\caption[]{\baselineskip=16pt This figure is a 
close-up, at two different contrast levels, of a
peculiar object 12\arcsec\ east of the quasar NAB~0205$+$02.  Stockton
and Mackenty (1987) first noticed it for being visible in their
[O~III] image, but not in the continuum.  The $HST$ image shown is
8\arcsec\ $\times$ 8\arcsec, and the
exposure time is 1900~sec. Because of its morphological peculiarity, we
refer to this object as the ``Barbell''.\protect\label{nabobj}}
%\end{figure}
%\nopagebreak
%\begin{figure}[h]

\caption[]{\baselineskip=16pt An expanded view of this spectacular merger of 
PG~1012$+$008,
which was ``caught in the act''.  The exposure time is
1400~sec.\protect\label{pg1012}}
%\end{figure}
%\nopagebreak
%\begin{figure}[h]
\caption[]{\baselineskip=16pt PSF subtracted image of the quasar PG~1309$+$355 
showing details in
the inner region of the host. Tightly wrapped spiral arms surround the 
nucleus.  The exposure time is
1400~sec.\protect\label{pg1309}}
%\end{figure}
%\nopagebreak
%\begin{figure}[h]
\caption[]{\baselineskip=16pt H~II regions in the spiral host of 
PG~1402$+$261. The H~II
regions listed in Table~\ref{hiitab} are identified in the upper panel.
The bar and probable inner ring are visible in the lower contrast image
reproduced in the lower panel.\protect\label{pg1402}}
%\end{figure}
%\nopagebreak
%\begin{figure}[h]
\caption[]{\baselineskip=16pt Absolute visual magnitude of 
the host galaxies 
versus the absolute visual magnitude
of the quasars.  The magnitudes and morphologies 
of the host galaxies are determined
using two-dimensional fits (cf. the summary in Table~\ref{summary}); the
absolute magnitudes of the quasars are taken from Table~\ref{journal}.
The diagonal line represents the average detection limit of smooth 
ellipticals in an idealized sample; see the
discussion in \S~\ref{subhostlum}.
For two of the quasars, PHL~909 and 3C~323.1,
the symbols overlap at $M_V({\rm host}) = -21.2$ and
$M_V({\rm QSO}) = -22.9$.
\protect\label{mvmv}}
\end{figure}

\newpage

\begin{deluxetable}{lcccccccclc}
\footnotesize
\tablewidth{7.1in}
\tablecaption{Quasar Sample\label{journal}}
\tablehead{
&&\colhead{Time}&\colhead{sky level}&&&&&\colhead{Radio}\\
\colhead{Object}&\colhead{Date}&\colhead{(s)}&
\colhead{(${\rm e^{-}~pix^{-1}~s^{-1}}$)}&
\colhead{$z$}&\colhead{${\rm kpc~arcsec^{-1}}$}
&\colhead{$V$}
&\colhead{$M_V$(QSO)$^{\rm a}$}&\colhead{Loud}&\colhead{FOS$^b$}}
\startdata
PG 0052$+$251 & 05 Dec 94& 1400 &0.114 & 0.155& 1.75& 15.4& $-$23.0& &   & \nl
PHL 909       & 17 Oct 94& 1400 &0.136 & 0.171& 1.88& 15.7& $-$22.9& &   & \nl
NAB 0205+02   & 26 Oct 94& 1400 &0.139 & 0.155& 1.75& 15.4& $-$23.0& &   & \nl
0316$-$346    & 20 Nov 94& 1400 &0.075 & 0.265& 2.55& 15.1& $-$24.5& &   & \nl
PG 0923$+$201 & 23 Mar 95& 1400 &0.136 & 0.190& 2.04& 15.8& $-$23.1& &   & \nl
\noalign{\medskip}			                
PG 0953$+$414 & 03 Feb 94& 1100 &0.110 & 0.239& 2.38& 15.3& $-$24.1& &K  & \nl
PKS 1004$+$130& 26 Feb 95& 1400 &0.150 & 0.240& 2.39& 15.2& $-$24.2&X&   & \nl
PG 1012$+$008 & 25 Feb 95& 1400 &0.129 & 0.185& 2.00& 15.6& $-$23.2& &   & \nl
HE 1029$-$140 & 06 Feb 95& 1400 &0.100 & 0.086& 1.22& 13.9& $-$23.2& &   & \nl
PG 1116$+$215 & 08 Feb 94& 1100 &0.128 & 0.177& 1.93& 14.7& $-$24.0& &K,O& \nl
\noalign{\medskip}			                
PG 1202$+$281 & 08 Feb 94& 1100 &0.116 & 0.165& 1.83& 15.6& $-$23.0& &K  & \nl
3C 273        & 05 Jun 94& 1100 &0.151 & 0.158& 1.78& 12.9& $-$25.6&X&K,O& \nl
PKS 1302$-$102& 09 Jun 94& 1100 &0.136 & 0.286& 2.67& 15.2& $-$24.6&X&K  & \nl
PG 1307$+$085 & 05 Apr 94& 1400 &0.129 & 0.155& 1.75& 15.1& $-$23.3& &   & \nl
PG 1309$+$355 & 26 Mar 95& 1400 &0.088 & 0.184& 1.99& 15.6& $-$23.2& &O  & \nl
\noalign{\medskip}			                
PG 1402$+$261 & 07 Mar 95& 1400 &0.089 & 0.164& 2.13& 15.5& $-$23.0& &   & \nl
PG 1444$+$407 & 27 Jun 94& 1100 &0.071 & 0.267& 2.56& 15.7& $-$23.9& &K,O& \nl
3C 323.1      & 09 Jun 94& 1100 &0.082 & 0.266& 2.55& 16.7& $-$22.9&X&   & \nl
PKS 2135$-$147& 15 Aug 94& 1400 &0.174 & 0.200& 2.11& 15.5& $-$23.5&X&O  & \nl
PKS 2349$-$014& 18 Sep 94& 1400 &0.161 & 0.173& 1.90& 15.3& $-$23.4&X&   & \nl
\enddata
\tablenotetext{a} { Computed for $\Omega_0 = 1.0$ and $H_0 = 100$
km s$^{-1}$Mpc$^{-1}$.  In this cosmology, brightest cluster galaxies
have $M_V~\approx~-22.0$ (Hoessel \& Schneider 1985; Postman \& Lauer
1995) and the characteristic (Schechter-) magnitude for field galaxies
is $M_V^*~=~-20.5$ (Schechter 1976; Kirshner et al.~1983; Efstathiou,
Ellis, \& Peterson 1988).  }
\tablenotetext{b} { K = $HST$ Quasar Absorption Line Key Project;  O = Other 
FOS observations. }
\end{deluxetable}

\begin{deluxetable}{lrcclc}
\tablecaption{Size and Morphology of Host Galaxies\label{size}}
\tablehead{
\colhead{Object}&\multicolumn{2}{l}{Major\ Diameter}
&\colhead{Isophote (F606W)}&\multicolumn{1}{l}{Morphology}\\
&\colhead{\arcsec}&\colhead{kpc}&\colhead{mag arcsec$^{-2}$}}
\startdata
PG 0052$+$251  & 20& 35& 25.1&  Sb &\nl
PHL 909        & 19& 36& 24.7&  E4 &\nl
NAB 0205$+$02  &  9& 16& 25.0&  S0? &\nl
0316$-$346     & 17& 43& 25.6&  Complex, interaction &\nl
PG 0923$+$201  & 13& 27& 24.8&  E1 &\nl
\noalign{\medskip}
PG 0953$+$414  & 11& 26& 25.8&  Faint, tail?  &\nl
PKS 1004$+$130 & 15& 36& 25.0&  E2 &\nl
PG 1012$+$008  & 18& 36& 25.0&  Interacting galaxies &\nl
HE 1029$-$140  & 40& 49& 26.0&  E1 &\nl
PG 1116$+$215  & 14& 27& 25.1&  E2 &\nl
\noalign{\medskip}
PG 1202$+$281  & 10& 18& 24.7&  E1, bright companion &\nl
3C 273         & 29& 52& 25.4&  E4 &\nl
PKS 1302$-$102 & 15& 40& 25.2&  E4 (?) two close companions &\nl
PG 1307$+$085  &  9& 16& 24.6&  Faint E1 (?) &\nl
PG 1309$+$355  & 18& 36& 25.7&  Sab &\nl
\noalign{\medskip}
PG 1402$+$261  & 14& 30& 25.6&  SBb &\nl
PG 1444$+$407  & 10& 26& 25.2&  E1 (?) &\nl
3C 323.1       & 11& 28& 25.1&  E3 (?) (bright companion) &\nl
PKS 2135$-$147  & 15& 32& 24.8&  E1 (companions) \nl
PKS 2349$-$014  & 21& 40& 24.5&  Complex, interacting &\nl
\enddata
\end{deluxetable}

\begin{deluxetable}{lrrrcccc}
\footnotesize
\tablecaption{Aperture Magnitudes for Host Galaxies\label{result}}
\tablehead{
&\colhead{inner}&\colhead{outer}
&\multicolumn{2}{c}{aperture}&
&\multicolumn{1}{c}{aperture}\\
&\colhead{radius}&\colhead{radius}
&\multicolumn{2}{c}{photometry}&
&\multicolumn{1}{c}{photometry}\\
\colhead{Object}&\colhead{(\arcsec)}&\colhead{(\arcsec)}
&\colhead{$m_{F606}$}&
\colhead{$M_{F606}$}&\colhead{$(F606-V)^a$}&\colhead{$M_{V}$}}
\startdata
PG 0052$+$251  &   1.0& 10.0&  17.1&  $-$21.3& $-$0.31& $-$21.0 \nl
PHL 909        &   1.0& 10.0&  17.5&  $-$21.1& $-$0.41& $-$20.7 \nl
NAB 0205$+$02  &   1.0&  4.5&  19.5&  $-$18.9& $-$0.31& $-$18.6 \nl
0316$-$346     &   1.0& 11.5&  18.2&  $-$21.4& $-$0.47& $-$20.9 \nl
PG 0923$+$201  &   1.0&  6.5&  18.3&  $-$20.6& $-$0.42& $-$20.2 \nl
\noalign{\medskip}             				   	  
PG 0953$+$414  &   1.0&  5.5&  19.1&  $-$20.3& $-$0.38& $-$19.9 \nl
PKS 1004$+$130 &   1.0&  7.5&  17.7&  $-$21.7& $-$0.48& $-$21.2 \nl
PG 1012$+$008  &   1.0&  4.5&  17.8&  $-$21.0& $-$0.39& $-$20.6 \nl
HE 1029$-$140  &   1.5& 20.0&  16.5&  $-$20.6& $-$0.35& $-$20.2 \nl
PG 1116$+$215  &   1.0&  8.0&  17.7&  $-$21.0& $-$0.41& $-$20.6 \nl
\noalign{\medskip}            				   	  
PG 1202$+$281  &   1.0&  5.0&  18.6&  $-$20.0& $-$0.40& $-$19.6 \nl
3C 273         &   2.0& 15.0&  16.6&  $-$21.9& $-$0.40& $-$21.5 \nl
PKS 1302$-$102 &   1.0&  7.5&  18.4&  $-$21.4& $-$0.50& $-$20.9 \nl
PG 1307$+$085  &   1.0&  4.5&  18.7&  $-$19.7& $-$0.37& $-$19.3 \nl
PG 1309$+$355  &   1.0&  9.0&  17.4&  $-$21.4& $-$0.35& $-$21.0 \nl
\noalign{\medskip}            				   	  
PG 1402$+$261  &   1.0&  7.5&  18.1&  $-$20.4& $-$0.30& $-$20.1 \nl
PG 1444$+$407  &   1.0&  5.0&  18.8&  $-$20.8& $-$0.47& $-$20.3 \nl
3C 323.1       &   1.0&  5.5&  18.9&  $-$20.7& $-$0.47& $-$20.2 \nl
PKS 2135$-$147 &   1.0&  7.5&  18.1&  $-$20.9& $-$0.45& $-$20.4 \nl
PKS 2349$-$014 &   1.0& 12.0&  16.7&  $-$22.0& $-$0.41& $-$21.6 \nl
\enddata
\tablenotetext{a} { Fukugita et al.\ (1995)}
\end{deluxetable}

\begin{deluxetable}{lrcccccccccc}
\footnotesize
\tablecaption{Model Fits to Stellar Quasar plus Host Galaxy \label{model}}
\tablehead{&\multicolumn{4}{c}{One$-$Dimensional}&&\multicolumn{4}{c}{Two$-$Dim
ensional}\\
&\multicolumn{2}{c}{GdV}&\multicolumn{2}{c}{Exp. Disk}&&\multicolumn{2}{c}{GdV}
&\multicolumn{2}{c}{Exp. Disk}\\
\colhead{Object}
&\colhead{$m_{F606}$}&\colhead{r(\arcsec)$^a$}&\colhead{$m_{F606}$}&\colhead{r(
\arcsec)$^b$}&
&\colhead{$m_{F606}$}&\colhead{r(\arcsec)$^a$}&\colhead{$m_{F606}$}&\colhead{r(
\arcsec)$^b$}}
\startdata
PG 0052$+$251  & 16.1& 4.7&  16.8& 1.4 &&  16.7 & 1.8 & 17.2 & 1.3   \nl
PHL 909        & 16.7& 2.5&  17.4& 1.0 &&  17.2 & 2.3 & 17.6 & 1.5   \nl
NAB 0205$+$02  & 18.0& 0.6&  18.7& 0.6 &&  18.4 & 0.7 & 19.0 & 0.7   \nl
0316$-$346     & 17.2& 3.4&  18.0& 1.1 &&  17.8 & 2.1 & 18.3 & 1.2   \nl
PG 0923$+$201  & 17.3& 2.5&  18.0& 1.0 &&  17.5 & 2.9 & 18.2 & 1.3   \nl
\noalign{\medskip}
PG 0953$+$414  & 17.9& 2.3&  18.5& 1.1 &&  18.2 & 1.8 & 18.8 & 1.1   \nl
PKS 1004$+$130 & 16.7& 1.6&  17.3& 0.9 &&  16.9 & 1.2 & 17.5 & 1.0   \nl
PG 1012$+$008  & 16.3& 6.2&  17.3& 1.4 &&  17.0 & 3.4 & 17.7 & 1.6   \nl
HE 1029$-$140  & 15.9& 2.8&  16.4& 1.5 &&  16.2 & 3.2 & 16.7 & 1.9   \nl
PG 1116$+$215  & 16.6& 1.9&  17.3& 1.0 &&  16.9 & 1.4 & 17.5 & 1.2   \nl
\noalign{\medskip}
PG 1202$+$281  & 17.4& 1.5&  18.1& 0.9 &&  17.7 & 1.4 & 18.3 & 1.0   \nl
3C 273         & 15.6& 2.3&  16.2& 1.3 &&  16.0 & 3.7 & 16.7 & 1.6   \nl
PKS 1302$-$102 & 17.1& 2.6&  17.8& 1.1 &&  17.7 & 1.4 & 18.2 & 1.1   \nl
PG 1307$+$085  & 17.4& 1.8&  18.1& 0.9 &&  17.8 & 1.3 & 18.4 & 1.0   \nl
PG 1309$+$355  & 16.4& 2.8&  17.1& 1.1 &&  16.8 & 2.0 & 17.3 & 1.2   \nl
\noalign{\medskip}
PG 1402$+$261  & 16.9& 2.2&  17.6& 1.0 &&  17.6 & 1.5 & 18.3 & 1.6   \nl
PG 1444$+$407  & 17.6& 1.3&  18.2& 1.0 &&  17.8 & 1.3 & 18.4 & 1.0   \nl
3C 323.1       & 17.8& 1.4&  18.4& 0.9 &&  18.1 & 1.6 & 18.7 & 1.0   \nl
PKS 2135$-$147 & 17.2& 2.0&  17.8& 1.1 &&  17.4 & 2.6 & 18.0 & 1.3   \nl
PKS 2349$-$014 & 15.9& 5.6&  16.6& 1.6 &&  16.2 & 4.8 & 16.8 & 2.5   \nl
\enddata
\tablenotetext{a} { Effective radius. }
\tablenotetext{b} { Exponential scale length. }
\end{deluxetable}

\begin{deluxetable}{lcccccc}
\footnotesize
\tablecaption{Absolute Visual Magnitudes for Quasar Host
 Galaxies\label{bestmodel}}
\tablehead{&\multicolumn{2}{c}{One$-$Dimensional}&
&\multicolumn{2}{c}{Two$-$Dimensional}\\
\colhead{Object}
&\colhead{$M_V$(1-D)}&\colhead{best~model}&&\colhead{$M_V$(2-D)}&
\colhead{best~model}}
\startdata
PG 0052$+$251    & $-$21.3  & Disk & &  $-$20.9 &  Disk  & \nl
PHL 909          & $-$21.5  & GdV & &   $-$21.0 &  GdV   & \nl
NAB 0205$+$02    & $-$19.4  & Disk & &  $-$19.1 &  Disk  & \nl
0316$-$346       & $-$21.1  & Disk & &  $-$20.8 &  Disk  & \nl
PG 0923$+$201    & $-$21.2  & GdV & &   $-$21.0 &  GdV   & \nl
\noalign{\medskip}
PG 0953$+$414    & $-$20.5  & Disk & &  $-$20.2 &  Disk  & \nl
PKS 1004$+$130   & $-$22.5  & GdV & &   $-$22.0 &  GdV   & \nl
PG 1012$+$008    & $-$22.1  & GdV & &   $-$20.7 &  Disk  & \nl
HE 1029$-$140    & $-$20.8  & GdV & &   $-$20.5 &  GdV   & \nl
PG 1116$+$215    & $-$21.7  & GdV & &   $-$21.4 &  GdV   & \nl
\noalign{\medskip}
PG 1202$+$281    & $-$20.8  & GdV & &   $-$20.5 &  GdV   & \nl
3C 273           & $-$22.5  & GdV & &   $-$22.1 &  GdV   & \nl
PKS 1302$-$102   & $-$21.5  & Disk & &  $-$21.1 &  Disk  & \nl
PG 1307$+$085    & $-$20.6  & GdV & &   $-$20.2 &  Disk  & \nl
PG 1309$+$355    & $-$21.3  & Disk & &  $-$21.1 &  Disk  & \nl
\noalign{\medskip}
PG 1402$+$261    & $-$20.6  & Disk & &  $-$19.9 &  Disk  & \nl
PG 1444$+$407    & $-$20.9  & Disk & &  $-$20.5 &  Disk  & \nl
3C 323.1         & $-$21.3  & GdV & &   $-$21.0 &  GdV   & \nl
PKS 2135$-$147   & $-$21.3  & GdV & &   $-$21.1 &  GdV   & \nl
PKS 2349$-$014   & $-$22.4  & GdV & &   $-$22.1 &  GdV   & \nl
\enddata
\end{deluxetable}

\begin{table*}[htb]
\centering
\begin{minipage}{5in}
\caption[]{Galaxy Companions Brighter than $M_{\rm (F606W)} =
- -16.5$\hfil\break
 within 25 kpc of the Quasar\label{comps}}
\begin{tabular*}{5in}{lcrrcrr}
\hline\hline
\noalign{\medskip}
\multicolumn{1}{c}{Quasar}&Number
of&\multicolumn{2}{r}{Distances}&&\multicolumn{2}{c}{Magnitudes}\\
&Companions&\multicolumn{1}{c}{ $''$}&
\multicolumn{1}{c}{\ kpc}&&\multicolumn{1}{r}{$m_{F606}$}
&\multicolumn{1}{c}{$\phantom{m}M_{F606}$}\\
\noalign{\medskip\hrule\medskip}
PG 0052$+$251    &     1   &    14.1 &  24.6  &&   18.8& $ -19.6$\\
PHL 909          &     1   &    12.5 &  23.5  &&   21.4& $ -17.2$\\
NAB 0205$+$02    &     1   &     8.3 &  14.5  &&   20.0& $ -18.4$\\
PG 0923$+$201    &     2   &    10.9 &  22.2  &&   19.5& $ -19.4$\\
                 &         &    11.0 &  22.5  &&   18.0& $ -20.9$\\
PG 0953$+$414    &     1   &     8.2 &  19.6  &&   22.7& $ -16.7$\\
PG 1012$+$00     &     2   &     3.3 &   6.7  &&   17.6& $ -21.2$\\
                 &         &     6.8 &  13.7  &&   19.0& $ -19.8$\\
PG 1116$+$215    &     1   &    12.3 &  23.8  &&   19.3& $ -19.4$\\
PG 1202$+$281    &     3   &     5.2 &   9.5  &&   18.9& $ -19.7$\\
                 &         &     8.4 &  15.3  &&   21.5& $ -17.1$\\
                 &         &     9.6 &  17.5  &&   21.3& $ -17.3$\\
PKS 1302$-$102   &     2   &     1.1 &   2.9  &&   20.3& $ -19.5$\\
                 &         &     2.3 &   6.2  &&   21.5& $ -18.3$\\
HE 1029$+140^{\rm a}$& 1   &     4.1 &   5.0  &&   20.7& $ -16.4$\\
3C 323.1         &     1   &     2.7 &   6.9  &&   20.6& $ -19.0$\\
PKS 2135$-$147   &     2   &     1.9 &   3.9  &&   19.5& $ -19.5$\\
                 &         &     5.5 &  11.7  &&   19.7& $ -19.3$\\
PKS 2349$-$014   &     2   &     1.9 &   3.5  &&   20.8& $ -17.9$\\
                 &         &    11.6 &  22.0  &&   21.4& $ -17.3$\\
\noalign{\medskip\hrule\medskip}
\noalign{\vbox{\hsize=5.0in\noindent
$^{\rm a}$\ Absolute magnitude of the close companion is 
0.1~mag fainter than the assumed limiting magnitude.\hfil\break}}
\end{tabular*}
\end{minipage}
\end{table*}

\begin{deluxetable}{lrcccccc}
\tablecaption{Comparison with results of V\'eron-Cetty \& Woltjer 
(1990)\label{veron}}
\tablehead{
\colhead{Object}&\colhead{$m^{a}_i$}&\colhead{$M^{a}_{V(i)}$}&
\colhead{$M^{b}_{F606}$}&\colhead{$M^{b}_{V(F606)}$}&\colhead{$M^{a}_V(model)$}
&\colhead{$M^{b}_{V(F606)}({\rm 2-D})$}\\[3pt]\cline{2-5}
\noalign{\smallskip}
&\multicolumn{4}{c}{12.5 kpc  ---  25 kpc}}
\startdata
PKS 1302$-$102 & 19.5& $-$20.3& $-$20.4& $-$20.0&  $-$22.0& $-$21.1& \nl
PKS 2135$-$147 & 18.9& $-$19.9& $-$19.9& $-$19.4&  $-$22.2& $-$21.1& \nl
PKS 2349$-$014 & 17.6& $-$20.7& $-$20.8& $-$20.4&  $-$22.5& $-$22.1& \nl
\enddata
\tablenotetext{a} { V\'eron-Cetty \& Woltjer (1990)}
\tablenotetext{b} { $HST$ results, this paper}
\end{deluxetable}

\begin{deluxetable}{lrrccc}
\footnotesize
\tablecaption{Annular Magnitudes. $F606$ magnitudes of host galaxies 
measured in an annulus between radii of 6~kpc and 12~kpc for
$H_0~=~100~ {\rm km~s^{-1}Mpc^{-1} }$, $\Omega_0~=~1.0$. (This radii
corresponds approximately to 12.5~kpc and 25~kpc for 
$H_0~=~50~ {\rm km~s^{-1}Mpc^{-1} }$, $\Omega_0~=~0.0$)\label{metric}}
\tablehead{
\colhead{Object}&\colhead{r$^a_1(\arcsec)$}&\colhead{r$^b_2(\arcsec)$}
&\colhead{$m_{F606}$}&\colhead{$M_{F606}$}&\colhead{$M_V$}}
\startdata
PG 0052$+$251  & 3.43 & 6.86 & 18.3&$-$20.1  &$-$ 19.8   \nl
PHL 909        & 3.19 & 6.38 & 18.8&$-$19.8  &$-$ 19.4   \nl
NAB 0205$+$02  & 3.43 & 6.86 & 20.7&$-$17.7  &$-$ 17.4   \nl
0316$-$346     & 2.80 & 4.71 & 19.8&$-$19.8  &$-$ 19.3   \nl
PG 0923$+$201  & 2.96 & 5.91 & 19.3&$-$19.6  &$-$ 19.2   \nl
\noalign{\medskip}
PG 0953$+$414  & 2.52 & 5.04 & 20.1&$-$19.3  &$-$ 18.9   \nl
PKS 1004$+$130 & 2.51 & 5.02 & 18.9&$-$20.5  &$-$ 20.0   \nl
PG 1012$+$008  & 3.00 & 6.00 & 18.1&$-$20.7  &$-$ 20.3   \nl
HE 1029$-$140  & 5.56 &11.11 & 18.0&$-$19.1  &$-$ 18.7   \nl
PG 1116$+$215  & 3.11 & 6.22 & 19.0&$-$19.7  &$-$ 19.3   \nl
\noalign{\medskip}
PG 1202$+$281  & 3.28 & 6.56 & 19.9&$-$18.8  &$-$ 18.4   \nl
3C 273         & 3.37 & 6.74 & 17.8&$-$20.7  &$-$ 20.3   \nl
PKS 1302$-$102 & 2.25 & 4.49 & 19.4&$-$20.4  &$-$ 19.9   \nl
PG 1307$+$085  & 3.43 & 6.86 & 19.9&$-$18.5  &$-$ 18.1   \nl
PG 1309$+$355  & 3.02 & 6.03 & 18.7&$-$20.1  &$-$ 19.7   \nl
\noalign{\medskip}
PG 1402$+$261  & 3.28 & 6.56 & 19.5&$-$19.0  &$-$ 18.7   \nl
PG 1444$+$407  & 2.34 & 4.69 & 19.7&$-$19.9  &$-$ 19.4   \nl
3C 323.1       & 2.35 & 4.71 & 19.9&$-$19.7  &$-$ 19.2   \nl
PKS 2135$-$147 & 2.84 & 5.69 & 19.1&$-$19.9  &$-$ 19.4   \nl
PKS 2349$-$014 & 3.16 & 6.32 & 17.9&$-$20.8  &$-$ 20.4   \nl
\enddata
\tablenotetext{a} {\ inner radius = 6~kpc}
\tablenotetext{b} {\ outer radius = 12~kpc}
\end{deluxetable}

\begin{deluxetable}{lcclccr}
\tablecaption{Comparison between absolute $V$ magnitudes for quasar host
galaxies expected from $K$-band and
from $HST-F606W$ measurements\label{dunlop}}
\tablehead{
\colhead{Object}&\colhead{$K_{gal}^a$}&\colhead{$M_K$}&\colhead{$(V-K)^{b}$}&
\colhead{$M_{V(K)}$}&\colhead{$M_{V(F606)}({\rm 2-D})^c$}&\colhead{$\Delta 
M_V$}}
\startdata
PG 0052$+$251   &  15.14 & $-$23.3 & 3.90 &   $-$19.4  & $-$20.9 &1.5 \nl
PHL 909         &  14.40 & $-$24.2 & 3.95 &   $-$20.3  & $-$21.0 &0.7 \nl
PG 0923$+$201   &  14.95 & $-$23.9 & 4.00 &   $-$19.9  & $-$21.0 &1.1 \nl
PG 0953$+$414   &  15.28 & $-$24.1 & 4.20 &   $-$19.9  & $-$20.2 &0.3 \nl
PKS 1004$+$13   &  15.12 & $-$24.3 & 4.20 &   $-$20.1  & $-$22.0 &1.9 \nl
PG 1012$+$00    &  13.94 & $-$24.9 & 4.00 &   $-$20.9  & $-$20.7 &$-$0.2 \nl
PKS 2135$-$147  &  14.75 & $-$24.2 & 4.10 &   $-$20.1  & $-$21.1 &1.0 \nl
PKS 2349$-$014  &  13.98 & $-$24.7 & 3.95 &   $-$20.8  & $-$22.1 &1.3 \nl
\enddata
\tablenotetext{a} { Dunlop et al.\ 1993. }
\tablenotetext{b} { $(V-K)$ for elliptical galaxy from Bruzual \& Charlot 
(1993). }
\tablenotetext{c} { This work (derived from two-dimensional galaxy
model fitting). }
\end{deluxetable}

\begin{deluxetable}{lrcccr}
\tablecaption{Comparison between absolute $V$ magnitudes for quasar host
galaxies expected from $H$-band and
from $HST$-F606W measurements\label{mcleod}}
\tablehead{
\colhead{Object}&\colhead{$H_{gal}^a$}&\colhead{$(V-H)_{normal}^b$}&\colhead{$M
_{V(H)}$}&
\colhead{$M_{V(F606)}$(2-D)$^c$}
&\colhead{$\Delta M_V$}}
\startdata
PG 0052$+$251   & 14.46 & 3.16 & $-$20.8&  $-$20.9& $ $0.1 \nl
PG 0923$+$201   & 14.86 & 3.22 & $-$20.8&  $-$21.0& $ $0.2 \nl
PG 0953$+$414   & 15.38 & 3.32 & $-$20.7&  $-$20.2& $-$0.5 \nl
PKS 1004$+$130  & 14.86 & 3.32 & $-$21.2&  $-$22.0& $ $0.8 \nl
PG 1012$+$008   & 14.02 & 3.21 & $-$21.6&  $-$20.7& $-$0.9 \nl
PG 1116$+$215   & 13.97 & 3.20 & $-$21.6&  $-$21.4& $-$0.2 \nl
PG 1202$+$281   & 15.07 & 3.18 & $-$20.3&  $-$20.5& $ $0.2 \nl
3C 273          & 13.01 & 3.17 & $-$22.3&  $-$22.1& $-$0.2 \nl
PKS 1302$-$102  & 14.79 & 3.47 & $-$21.5&  $-$21.1& $-$0.4 \nl
PG 1307$+$085   & 15.24 & 3.16 & $-$20.0&  $-$20.2& $ $0.2 \nl
PG 1309$+$355   & 14.55 & 3.21 & $-$21.0&  $-$21.1& $ $0.1 \nl
PG 1402$+$261   & 14.95 & 3.18 & $-$20.4&  $-$19.9& $-$0.5 \nl
PG 1444$+$407   & 15.19 & 3.42 & $-$21.0&  $-$20.5& $-$0.5 \nl
3C 323.1        & 14.80 & 3.42 & $-$21.4&  $-$21.0& $-$0.4 \nl
\enddata
\tablenotetext{a} { McLeod \& Rieke (1994b). }
\tablenotetext{b} { McLeod \& Rieke (1995). }
\tablenotetext{c} { This work (derived from two-dimensional galaxy
model fitting.)}
\end{deluxetable}

\begin{deluxetable}{ccccc}
\tablewidth{11cm}
\tablecaption{H~II Regions in the Host Galaxy of PG~1402$+$261\
\label{hiitab}}
\tablehead{
\colhead{Region}&\colhead{$m_{F606W}$}&\colhead{$d$}&\colhead{$\Delta\alpha$}
&\colhead{$\Delta\delta$}\\
&&\colhead{($\arcsec$)}&\colhead{($\arcsec$)}&\colhead{($\arcsec$)}
}
\startdata
a&23.7&2.8&\llap{$-$}0.3&          2.8 \nl
b&24.3&2.9&\llap{$-$}1.5&          2.5 \nl
c&25.3&2.9&\llap{$-$}2.1&          2.0 \nl
d&25.6&3.5&\llap{$-$}3.3&          1.3 \nl
e&25.5&4.6&\llap{$-$}4.5&\llap{$-$}0.9 \nl
f&26.2&4.8&\llap{$-$}4.5&\llap{$-$}1.8 \nl
g&26.2&5.0&\llap{$-$}4.2&\llap{$-$}2.8 \nl
h&25.9&5.3&\llap{$-$}3.8&\llap{$-$}3.7 \nl
i&25.6&4.1&\llap{$-$}1.1&          4.0 \nl
j&25.2&4.0&\llap{$-$}2.7&          3.0 \nl
\enddata
\end{deluxetable}

\begin{deluxetable}{lcccclc}
\footnotesize
\tablecaption{Summary of Magnitudes and Morphology for Quasar Host
 Galaxies\label{summary}}
\tablehead{&&&\multicolumn{2}{c}{Two$-$Dimensional}\\
\colhead{Object}&\colhead{$z$}&\colhead{$m_{606}$(2-D)}&\colhead{$M_V$(2-D)}&\c
olhead{$r(\arcsec)^{a}$}&\colhead{Morphology}}
\startdata
PG 0052$+$251    & 0.155& 17.2 & $-$20.9 & 1.3 &  Sb & \nl
PHL 909          & 0.171& 17.2 & $-$21.0 & 2.3 &  E4 & \nl
NAB 0205$+$02    & 0.155& 19.0 & $-$19.1 & 0.7 &  S0? & \nl
0316$-$346       & 0.265& 18.3 & $-$20.8 & 1.2 &  Inter. & \nl
PG 0923$+$201    & 0.190& 17.5 & $-$21.0 & 2.9 &  E1 & \nl
\noalign{\medskip}	            	       	    
PG 0953$+$414    & 0.239& 18.8 & $-$20.2 & 1.1 &  ? & \nl
PKS 1004$+$130   & 0.240& 16.9 & $-$22.0 & 1.2 &  E2 & \nl
PG 1012$+$008    & 0.185& 17.7 & $-$20.7 & 1.6 &  Inter. & \nl
HE 1029$-$140    & 0.086& 16.2 & $-$20.5 & 3.2 &  E1 & \nl
PG 1116$+$215    & 0.177& 16.9 & $-$21.4 & 1.4 &  E2 & \nl
\noalign{\medskip}	            	       	    
PG 1202$+$281    & 0.165& 17.7 & $-$20.5 & 1.4 &  E1 & \nl
3C 273           & 0.158& 16.0 & $-$22.1 & 3.7 &  E4 & \nl
PKS 1302$-$102   & 0.286& 18.2 & $-$21.1 & 1.1 &  E4? & \nl
PG 1307$+$085    & 0.155& 17.8 & $-$20.2 & 1.3 &  E1? & \nl
PG 1309$+$355    & 0.184& 17.3 & $-$21.1 & 1.2 &  Sab & \nl
\noalign{\medskip}	            	       	    
PG 1402$+$261    & 0.164& 18.3 & $-$19.9 & 1.6 &  SBb & \nl
PG 1444$+$407    & 0.267& 18.4 & $-$20.5 & 1.0 &  E1? & \nl
3C 323.1         & 0.266& 18.1 & $-$21.0 & 1.6 &  E3? & \nl
PKS 2135$-$147   & 0.200& 17.4 & $-$21.1 & 2.6 &  E1 & \nl
PKS 2349$-$014   & 0.173& 16.2 & $-$22.1 & 4.8 &  Inter. & \nl
\enddata
\tablenotetext{a} {Effective radius or exponential scale length. }
\end{deluxetable}
\end{document}